\documentclass[%
 reprint,
superscriptaddress,
 amsmath,amssymb,
 aps, prb,
twocolumn]{revtex4-2}
\usepackage{latexsym}
\usepackage{graphicx}
\usepackage{verbatim}
\usepackage{multirow}
\usepackage [english]{babel}
\usepackage [autostyle, english = american]{csquotes}
\MakeOuterQuote{"}
\usepackage{amsmath}
\usepackage{mathrsfs}
\usepackage{float}
\usepackage{caption}
\usepackage{bm}
\usepackage{mathtools}
\usepackage[usenames, dvipsnames]{color}
\usepackage{tabularx}
\usepackage{amsfonts}
\DeclareMathSymbol{\shortminus}{\mathbin}{AMSa}{"39}
\usepackage{relsize}
\usepackage{subcaption}
\usepackage{mathtools}
\usepackage{slashed}
\usepackage{physics}	
\usepackage{graphicx}   
\usepackage{epstopdf}
\usepackage{hyperref}   
\usepackage{bbold}
\usepackage{wasysym}
\usepackage{amsmath}
\usepackage{bm}
\usepackage{pdfsync}
\begin{document}

\title{Theory of Half-Integer Fractional Quantum Spin Hall Insulator Edges }
\author{Julian May-Mann} 
\email{maymann@stanford.edu}
\affiliation{Department of Physics, Stanford University, Stanford, CA 94305, USA}
\author{Ady Stern}
\affiliation{Department of Condensed Matter Physics, Weizmann Institute of Science, Rehovot 76100, Israel}
\author{Trithep Devakul} 
\affiliation{Department of Physics, Stanford University, Stanford, CA 94305, USA}

\begin{abstract}
We study the edges of fractional quantum spin Hall insulators (FQSH) with half-integer spin Hall conductance. These states can be viewed as symmetric combinations of a spin-up and spin-down half-integer fractional quantum Hall state (FQH) that are time-reversal invariant, and conserve the z-component of spin. We consider the non-Abelian states based on the Pfaffian, anti-Pfaffian, PH-Pfaffian, and 221 FQH, and generic Abelian FQH. For strong enough spin-conserving interactions, we find that all the non-Abelian and Abelian edges flow to the same fixed point that consists of a single pair of charged counter-propagating bosonic modes. If spin-conservation is broken, the Abelian edge can be fully gapped in a time-reversal symmetric fashion. The non-Abelian edge with broken spin-conservation remains gapless due to time-reversal symmetry, and can flow to a new fixed point with a helical gapless pair of Majorana fermions. We discuss the possible relevance of our results to the recent observation of a half-integer edge conductance in twisted MoTe$_2$. 
\end{abstract}
\maketitle

\textit{Introduction.}--- 
Fractionalization is one of the most fascinating consequences of many-body correlations and entanglement in condensed matter physics. 
The prime example is the fractional quantum Hall (FQH) effect, in which an electron is effectively split into multiple emergent quasiparticle excitations, known as anyons, with unusual exchange and braiding statistics\cite{stormer1999fractional, stern2008anyons, cage2012quantum}.
Anyons with non-Abelian statistics are particularly sought after for their potential applications in topological quantum computing\cite{sarma2005topologically, nayak2008non}. 
Recently, two-dimensional moiré materials have emerged as an ideal platform for studying the interplay of symmetry, fractionalization, and correlations\cite{andrei2020graphene, balents2020superconductivity,he2021moire, mak2022semiconductor}.
Notably, the FQH effect at zero magnetic field,\cite{regnault2011fractional,neupert2011fractional, tang2011high, sheng2011fractional, wu2012zoology}, has beenobserved in twisted bilayer MoTe$_2$ (tMoTe$_2$) \cite{ cai2023signatures,park2023observation, zeng2023thermodynamic, xu2023observation}and in substrate-aligned rhombohedral pentalayer graphene\cite{lu2024fractional}.

In addition to time-reversal symmetry (TRS) breaking states, like the FQH, there are also TRS invariant fractionalized states, known as fractional topological insulators (FTI) or fractional quantum spin Hall (FQSH) insulators~\cite{FTI_review}.  
Very recently, experimental evidence for such a state was observed in tMoTe$_2$ in the form of a quantized two-terminal conductance of $\sigma_2=\frac{3}{2}\frac{e^2}{h}$ per edge at a filling of $n=3$ holes per moir\'e unit cell~\cite{kang2024observation}. The Hall conductance also vanishes at this filling.   
Remarkably, these observations are consistent with a spin-up $\sigma_{xy}=\frac{3}{2}$ FQH state in parallel with its spin-down time-reversed partner, resulting in a
 FQSH state with spin-Hall conductivity $\sigma_{sH}=\frac{3}{2}$.
However, the fate of the counter-propagating edge modes in the presence of interactions between the two spin species is subtle\cite{levin2009fractional, FTI_review}.

Motivated by this, we consider FQSH states with half-integer spin-Hall conductances.
Specifically, these are topologically ordered systems with U$(1)$ charge conservation, U$(1)_{\text{s}}$ $s^z$-spin conservation, and TRS.
This choice of symmetries is informed by those of the effective degrees of freedom in tMoTe$_2$\cite{xiao2012coupled, wu2019topological, devakul2021magic}, although our analysis is entirely general.
The topological order of a half-integer FQSH state can be either Abelian or non-Abelian.
While Abelian states have been considered in detail\cite{bernevig2006quantum, levin2009fractional,qi2011generic,levin2011exactly, lu2012symmetry,Levin_2012,FTI_review, neupert2015fractional}, there have been fewer works on the non-Abelian states~\cite{cappelli2013partition, cappelli2015stability}.
The possibility of a half-integer FQSH state invites further exploration into the non-Abelian states, as there is overwhelming evidence that the half-integer FQH states in GaAs\cite{willett1987observation} are non-Abelian\cite{willett2013quantum, morf1998transition,  lu2010superconducting, storni2010fractional, pakrouski2015phase, tylan2015phase, banerjee2017observed, banerjee2018observation,  ma2022fractional}. 

In this work, we are primarily interested in the edge structure of the half-integer FQSH states, as the edge modes are directly probed in transport-based experiments.
There are several possible Abelian and non-Abelian half-integer FQSH states, which have different bosonic and/or Majorana edge modes when the spin-up and spin-down degrees of freedom are fully decoupled from each other.
Despite this, we find that when $s_z$-conserving interactions are included all the Abelian and non-Abelian FQSH edges we consider here have a shared ``minimal'' form.
This minimal edge has a single non-chiral pair of charged bosons, one per spin.
Depending on the details of the interactions, the decoupled edge can have a weak coupling instability towards forming the minimal edge. 
This indicates that transport alone may not be sufficient to differentiate Abelian and non-Abelian FQSHs.

While the half-intger Abelian and non-Abelian FQSHs have a shared minimal edge theory when $s_z$ conservation is preserved, this is not true if we consider breaking $s^z$-conservation at the edge but preserving TRS. 
Previous works have shown that the Abelian edge can be fully gapped~\cite{levin2009fractional}. 
In contrast, we show that the non-Abelian edges remain gapless as long as TRS is preserved. 

It will be useful to distinguish between a ``quantum spin Hall insulator'' and a ``topological insulator'', as the terms have been used synonymously is the literature. Here, we shall use ``quantum spin Hall insulators'' for insulators with gapless edge modes that are stable in presence of charge conservation, $s^z$ conservation, and time-reversal symmetry. We shall use ``topological insulators'' for insulators with gapless edge modes that need only charge conservation and time-reversal symmetry to be protected from gapping. Since the half-integer non-Abelian FQSHs with broken $s^z$-conservation have protected gapless edges, they are non-Abelian FTIs~\cite{levin2009fractional}.

\textit{Effective theories of the interacting edge.}--- The first main result of this work is that the edges of all half-integer FQSHs we consider in this work can be gapped up to a single gapless pair of charged bosonic modes with the effective Lagrangian
\begin{equation}\begin{split}
\mathcal{L}_{{\text{\smaller{FQSH}}}} &= \frac{\rho_{sH}}{4\pi} \partial_{x}\phi_{R} (\partial_t  - v \partial_x) \phi_{R}  \\ &+ \frac{\rho_{sH}}{4\pi} \partial_{x}\phi_{L} (-\partial_t  - v \partial_x) \phi_{L} - u \partial_x \phi_{R} \partial_x \phi_{L}\\
&+ \frac{1}{2\pi}  \epsilon^{\mu\nu} [\partial_\mu (\phi_{R} + \phi_{L}) A_\nu + \partial_\mu  (\phi_{R} - \phi_{L})  A^{\text{s}}_\nu].
\label{eq:effectiveLag}\end{split}\end{equation}
Here $\phi_R$ and $\phi_L$ are chiral (right moving) and anti-chiral (left moving) compact bosons with radius $2\pi$, and $\rho_{sH} = \sigma_{sH}^{-1}$ is the spin Hall resistance, which is detemined by the bulk topological order. We work in units where $e = \hbar = 1$, except when dimensionful quantities are relevant. The bosons have the same charge but opposite spin, as evident from the couplings to charge and spin probe gauge fields, $A$, and $A^{\text{s}}$. Under time-reversal symmetry, $\phi_{R}\rightarrow \phi_L$ and $\phi_{L}\rightarrow \phi_R + \pi \sigma_{sH}$.

Since a FQSH must have at least one pair of bosons to carry the charge and spin, Eq.~\ref{eq:effectiveLag} is the \textit{minimal} edge theory of a FQSH, i.e., the chiral and anti-chiral sectors have the lowest possible central charges\cite{francesco2012conformal}. Eq.~\ref{eq:effectiveLag} is stable due to charge and $s^z$-conservation. To show this, consider the dual bosons, $\varphi = \phi_R + \phi_L$ and $\theta = \phi_R - \phi_L$. Gapping the edge amounts to condensing, $\varphi$ or $\theta$. However, since $\varphi$ carries spin and $\theta$ carries charge, gapping the edge necessarily involves breaking charge or $s^z$-conserving. Small TRS breaking effects that do not break charge or $s_z$ conservation amount to a relative change of $v$ for $\phi_{R}$ and $\phi_L$, and are therefore inconsequential.

Although the FQSH is a fermionic system, the minimal edge does not have any local fermionic operators. Rather, all local operators are bosons, indicating that there is a gap for adding a single electron (see Ref.~\cite{cano2014bulk} for a similar effect in quantum Hall systems). The local bosons with the lowest (non-zero) charges are $e^{i 4 \phi_{R}}$, $e^{-i 4 \phi_{L}}$, both of which carry even charge $4\sigma_{sH}$.

If we consider breaking $s^z$ conservation of a non-Abelian half-integer FQSH but preserving TRS (for example, by adding Rashba spin-orbit coupling), we find that the edge remains gapless. The state with broken $s^z$-conservation is therefore a non-Abelian FTI. However, when $s^z$-conservation is broken, the bosonic edge in Eq.~\ref{eq:effectiveLag} is no longer the minimal edge of the non-Abelian FTI. Rather, the minimal edge consists of a single non-chiral pair of Majorana fermions, 
\begin{equation}
    \mathcal{L}_{\text{\smaller{FTI}}} = i\psi_{R} (\partial_t + v_f \partial_x ) \psi_{R} + i \psi_{L} (\partial_t \shortminus v_f\partial_x ) \psi_{L}
\label{eq:effectiveMagLag}\end{equation}
This is the second main result of this work. We establish Eq.~\ref{eq:effectiveMagLag} by noting that the non-Abelian FTI edge can be formulated in terms of a single non-chiral pair of charged bosons, and an odd number of non-chiral pairs of Majoranas.
The charged bosons and an even number of Majorana pairs can be gapped out while preserving charge conservation and TRS. This leaves a single gapless pair of Majoranas that cannot be gapped out without breaking TRS. 
Since Eq.~\ref{eq:effectiveMagLag} only has gapless neutral excitations, the FTI can be an electrical insulator that may conduct heat on its edges.

Before the technical discussion showing these results, it will be useful to discuss the experimentally relevant properties of the minimal FQSHI and FTI edges. For the minimal FQSH edge the minimal charge that can be injected into the edge at low energy  is $4\sigma_{sH}e$, instead of $e$. This difference will manifest in tunneling measurements from an outside source into the edge, and in particular in shot noise measurements. Next, we consider the electric conductance of the minimal FQSH edge, $\sigma_2$. To model a real setup, we consider a long section of the 1D edge in Eq.~\ref{eq:effectiveLag} that is connected to two terminals. Since translation symmetry is broken in this geometry, the Luttinger parameter, $\kappa$, and velocity $v$ can be functions of position.
The conductance between two terminals can be calculated with the Kubo formula. Following the results of Ref.~\onlinecite{maslov1995landauer}, $\sigma_2 = \sigma_{sH}\frac{e^2}{h} \kappa(x_{\text{t}})$, where $\kappa(x_{\text{t}})$ is the value of the Luttinger parameter at terminals. For Fermi-liquids, the low-energy spin-up and spin-down degrees of freedom that couple to the FQSH edge should be fully decoupled from one another. This corresponds to $\kappa(x_{\text{t}}) = 1$, leading to a two-terminal conductance of $\sigma_2 = \sigma_{sH}\frac{e^2}{h}$. 

We also consider the thermal conductance of the FQSH edge. For a system of length $L$ that is much larger than the thermal relaxation length $\xi$, the thermal conductance of a non-chiral edge scales like $\tfrac{ |c|_{\text{tot}}}{1 + L/\xi} \kappa_0 T$, where $|c|_{\text{tot}}$ is sum of the absolute values of the central charges of all gapless edge modes, and $\kappa_0 = \tfrac{\pi^2 k_B^2}{3h}$ is the thermal conductance quanta\cite{banerjee2017observed}. Clearly, the thermal conductance vanishes as $L \rightarrow \infty$. However, in the opposite limit $L \rightarrow 0$ (obtained by extrapolating backwards while keeping $L\gg\xi$), the thermal conductance approaches $|c|_{\text{tot}}\kappa_0 T$.  For the bosonic edge, $|c|_{\text{tot}} = 2$, such that the $L \rightarrow 0$ limit is $2\kappa_0 T$. The minimal FTI also has thermal transport signatures due to the presence of the gapless Majorana fermions. Specifically, for minimal FTI edge the $L \rightarrow 0$ limit of the thermal conductance is $\kappa_0 T$.

\textit{Generalized-Pfaffian FQSHs.}--- To begin, we consider FQSHs with $\sigma_{sH} = \frac{1}{2}$. This state can be realized by combining a spin-up $\nu = \frac{1}{2}$ FQH and its spin-down counterpart. The FQH can be either Abelian or non-Abelian. We present a general discussion of the Abelain cases in Appendix~\ref{app:AbelainIntEdge}. The non-Abelian case is more interesting and relevant, and will be the focus of the rest of this work. Consider the non-Abelian Pfaffian (Pf)\cite{moore1991nonabelions}, PH-Pfaffian (PH-Pf)\cite{son2015composite}, 221\cite{jain1989incompressible}, and anti-Pfaffian (aPf)\cite{levin2007particle, lee2007particle} FQHs. The edges of these states all have a chiral charged boson accompanied by a chiral Majorana in the Pf state; an anti-chiral Majorana in the PH-Pf state; three chiral Majoranas in the 221 state; and three anti-chiral Majoranas in the aPf state. Motivated by this, we consider a class of FQSH made out of generalized-Pfaffian (gPf) states with $M$ Majoranas. We shall show that if double spin-flip interactions are included, the minimal edge theory in Eq.~\ref{eq:effectiveLag} with $\sigma_{sH} = \frac{1}{2}$ is a strong coupling fixed point of the edge. This is generically true for $M \geq 2$, and also true for $M = 1$ if we allow for  edge reconstruction. In Appendix~\ref{app:RGFlows} we further show that the gPF FQSH edges have a weak coupling instability towards the minimal edge theory in Eq.~\ref{eq:effectiveLag}, provided there are significantly strong repulsive density-density interactions between the up and down spins.

The Lagrangian for the gPf FQSH can be written in terms of fermionic and bosonic contributions,
\begin{equation}\begin{split}
\mathcal{L}_{f,m} =& i\psi_{\uparrow m} (\partial_t + v_f \partial_x ) \psi_{\uparrow m} + i \psi_{\downarrow m} (\partial_t \shortminus v_f\partial_x ) \psi_{\downarrow m},\\[5pt]
\mathcal{L}_b = &\tfrac{1}{4\pi} \partial_t \bm{\Phi}^T \bm{K}  \partial_x  \bm{\Phi} - \tfrac{1}{4\pi} \partial_x  \bm{\Phi}^T \bm{V} \partial_x \bm{\Phi} \\[1pt]
&+ \tfrac{1}{2\pi} \epsilon^{\mu\nu} \partial_\mu (\bm{t}\cdot \bm{\Phi}) A_\nu + \tfrac{1}{2\pi} \epsilon^{\mu\nu} \partial_\mu (\bm{s}\cdot \bm{\Phi})  A^{\text{s}}_\nu.
\label{eq:FermiAndBosonLagrangian}\end{split}\end{equation}
$\mathcal{L}_{f,m}$ is the Lagrangian for the $\psi_{\uparrow m}$, $\psi_{\downarrow m}$ Majoranas. We define $v_f$ such that the Majoranas co-propagate (counter propagate) with the corresponding bosons when it is positive (negative). 
The bosonic Lagrangian  $\mathcal{L}_b$ is written in the  K-matrix form\cite{wen2004quantum}, where $\bm{K} = \text{diag}(2,-2)$, $\bm{\Phi} = (\phi_{P\uparrow}, \phi_{P\downarrow})$, and $\bm{V}$ encodes the velocity and Luttinger parameter of the bosons. The charge vector $\bm{t} = (1,1)$ and spin vector $\bm{s} = (\frac{1}{2},-\frac{1}{2})$ so the bosons carry the same charge and opposite spin. We will focus on the case where $M$ is odd, as the even $M$ gPF FQSHs are Abelian. 

Local operators can be written in terms of the following $2M$ electron operators  $\chi^\dagger_{\uparrow m} = \psi_{\uparrow m} e^{i 2 \phi_{\uparrow P}}$, and $\chi^\dagger_{\downarrow m} = \psi_{\downarrow m} e^{-i 2 \phi_{\downarrow}}$\footnote{Here we are leaving the Klein factors needed for the fermion anti-commutation relationships implicit}. There are other local charge $e$ fermions in the theory that are combinations of the above $2M$ operators. These electron operators are invariant under the following $2M$ parity transformations: $\psi_{\uparrow/\downarrow m} \rightarrow -\psi_{\uparrow/\downarrow m}$, $\phi_{ \uparrow/\downarrow P} \rightarrow \phi_{\uparrow/\downarrow P} + \frac{\pi}{2}$ (for each $m$, $\uparrow$, and $\downarrow$ individually)\cite{milovanovic1996edge, sohal2020entanglement, lim2021disentangling}. Any local operator must also be invariant under this transformation. Additionally, since any physical state can be expressed in terms of local operators operating on the vacuum, these transformations indicate that the physical Hilbert space is smaller than the expanded Hilbert space spanned by the Majorana and bosonic sectors. Notably, the physical Hilbert space only contains sectors where the Majorana fermion parity is equal to the electric charge parity for each spin\footnote{We will only consider systems where the bulk topological order is defined on a manifold with a single edge, (e.g., discs). If the manifold has multiple decoupled edges, (e.g., a corbino disk or a cylinder) only the net parity of all edges is conserved}. Here, we work in the expanded Hilbert space unless otherwise specified, and implicitly project onto the physical Hilbert space at the end. It is worth explicitly stating that particles that are gapped (gapless) in the expanded Hilbert space remain gapped (gapless) after projection. 

A caveat of the existence of the parity transformations is that the TRS can be equivalently thought of as acting non-trivially on the bosons or the Majoranas. The bosonic TRS is $\psi_{\uparrow m} \leftrightarrow \psi_{\downarrow m}$, $\phi_{ \uparrow P} \rightarrow \phi_{\downarrow P}$ and $\phi_{\downarrow P} \rightarrow \phi_{\uparrow P} + \pi/2$. The Majorana TRS is $\psi_{\uparrow m} \rightarrow \psi_{\downarrow m}$, $\psi_{\downarrow m} \rightarrow -\psi_{\uparrow m}$ and $\phi_{\downarrow P} \leftrightarrow \phi_{\uparrow P}$. These two definitions are equivalent up to a parity transformation, and therefore have the same action on the physical Hilbert space. Local operators that break (do not break) TRS are necessarily odd (even) under both the bosonic and Majorana TRS. Operators that are odd under only one of the TRS definitions are not invariant under the parity transformation and are therefore non-local. Note that TRS is only broken if a \textit{local} time-reversal odd operator has an expectation value.

We now consider adding interactions to Eq.~\ref{eq:FermiAndBosonLagrangian}. For  $M \geq 3$, the following double spin-flip interaction induces a Majorana gap,
\begin{equation}\begin{split}
\frac{J}{2} \text{$\sum_{m, m'}$} \chi^\dagger_{\uparrow m} \chi_{\uparrow m'} \chi^\dagger_{\downarrow m'} \chi_{\downarrow m} = -\frac{J}{2}(\sum_{m} i\psi_{\uparrow m} \psi_{\downarrow m})^2.
\label{eq:GrossNeveu}\end{split}\end{equation}
This interaction is equivalent to that of the $O(M)$ Gross-Neveu model\cite{gross1974dynamical}. In the strong coupling limit of the $O(M)$ Gross-Neveu model with $M \geq 3$, $\langle i \psi_{\uparrow m} \psi_{\downarrow m}\rangle \neq 0$, which gaps out the Majoranas\cite{witten1978some, zamolodchikov1978exact, fendley2001bps}. This can also be confirmed by turning $M-1$ of the Majoranas into $\frac{M-1}{2}$ complex fermions, and then bosonizing the complex fermions. The resulting interaction has the same form as those we consider in Appendix~\ref{app:MajAndVertex}. Eq.~\ref{eq:GrossNeveu} leads to an edge that has gapped Majorana modes, while conserving both charge and spin. TRS is also unborken when $\langle i \psi_{\uparrow m} \psi_{\downarrow m} \rangle \neq 0$, as the Majorana bilinear is odd under the parity transformation and is therefore a non-local operator. The fact that TRS is unbroken can also be confirmed by explicitly considering the action of TRS on the projected states in the physical Hilbert space. The remaining gapless fields are described by Eq.~\ref{eq:effectiveLag} with $\phi_{R} = \phi_{\uparrow P}$, and $\phi_{L} = \phi_{\downarrow P}$.

For $M = 1$ the only $s_z$ preserving interaction we can add to the gPF edge is $\chi^\dagger_{\uparrow 1 }\chi_{\uparrow 1 } \chi^\dagger_{\downarrow 1 }\chi_{\downarrow 1 }$, which simply changes the Luttinger parameter of bosons. However, it is possible to gap out the Majoranas with symmetry preserving interaction if we consider edge reconstruction effects. Here, edge reconstruction amounts to adding a purely 1+1D quantum wire  (1DQW) to the edge. The bosonized fermionic modes of the 1DQW are $\Psi^\dagger_{\uparrow 1 }= e^{i \phi_{\uparrow 1}}$, $\Psi^\dagger_{\uparrow 2 }=e^{-i \phi_{\uparrow 2}}$, and their time-reversed counterparts. Here $\Psi_{\uparrow 1 }$ and $\Psi_{\downarrow 2 }$ are right moving fermions, while $\Psi_{\uparrow 2 }$ and $\Psi_{\downarrow 1 }$ are left moving. Using the 1DQW fermions, we have the double spin flip interactions
\begin{equation}\begin{split}
&\frac{J'}{2}\chi^\dagger_{\uparrow 1} \chi_{\downarrow 1} [\Psi^\dagger_{\downarrow 1 } \Psi_{\uparrow 1 } + \Psi^\dagger_{\downarrow 2 }\Psi_{\uparrow 2 }] + h.c.\\
& = J' i \psi_{\uparrow 1} \psi_{\downarrow 1} [\cos(2\varphi_{P} -\varphi_{1}) + \cos(2\varphi_{P} - \varphi_{2})].
\label{eq:1MgapTerm}\end{split}\end{equation}
where we have introduced the non-chiral boson, $\varphi_{\alpha} = \phi_{\uparrow \alpha} + \phi_{\downarrow \alpha}$ with dual boson $\theta_{\alpha} = \phi_{\uparrow \alpha} - \phi_{\downarrow \alpha}$, for $\alpha = P$, $1$, $2$. In Appendix~\ref{app:MajAndVertex} we show that in the strong coupling limit of Eq.~\ref{eq:1MgapTerm}, the Majorana bilinear and cosine terms both acquire expectation values, gapping out the Majorana fermions, and four of the six bosonic fields. Again, this does not break TRS, as the cosine terms and Majorana bilinear are non-local. The remaining gapless bosons realize the minimal edge in Eq.~\ref{eq:effectiveLag} with $\phi_{R} = \phi_{\uparrow P} + \phi_{\uparrow 1} + \phi_{\uparrow 2}$ and $\phi_{L} = \phi_{\downarrow P} + \phi_{\downarrow 1} + \phi_{\downarrow 2}$.

\textit{Integer+Generalized-Pfaffian FQSHs.}--- We now extend our arguments to other non-Abelian half-integer FQSH. We consider a state where the gPf FQSH modes are in parallel with $N$ integer quantum spin Hall modes, leading to a spin Hall conductance of $\sigma_{sH} = N+\frac{1}{2}$. 
The edge Lagrangian is $\mathcal{L} = \sum_m \mathcal{L}_{f,m} + \mathcal{L}_b,$.  The fermionic part  $\mathcal{L}_{f,m}$ is identical to that in Eq.~\ref{eq:FermiAndBosonLagrangian}. The bosonic part has now $2N+2$ bosonic fields $\bm{\Phi} = (\phi_{P\uparrow}, \phi_{\uparrow 1}..., \phi_{\uparrow N}, \phi_{P\downarrow}, \phi_{\downarrow 1}...,\phi_{\downarrow N})$, with K-matrix $\bm{K} = \text{diag}(2,1,...,-2,-1,...)$, and charge and spin vectors $\bm{t} = (1...1)$ and $\bm{s} = (\frac{1}{2},...,-\frac{1}{2}...)$. The local electron operators are $\chi^\dagger_{\uparrow m}$, $\chi^\dagger_{\downarrow m}$,  $\Psi^\dagger_{\uparrow n} = e^{i \phi_{\uparrow n}}$, and $\Psi^\dagger_{\downarrow n} = e^{-i \phi_{\downarrow n}}$.

This integer+gPf FQSH is driven to the interacting edge (Eq.~\ref{eq:effectiveLag}) by the following double spin-flip operations, in which the gPf state is coupled to each of the $N$ pairs of counter-propagating modes
\begin{equation}\begin{split}
\frac{J''}{2}&\sum\limits_{mn}\chi^\dagger_{\uparrow m}\chi_{ \downarrow m}\Psi^\dagger_{\downarrow n}\Psi_{\uparrow n} 
= J''\sum\limits_{mn} i \psi_{\uparrow m}\psi_{\downarrow m} \cos(2\varphi_{P} - \varphi_{n})
\label{eq:DoubleSpinFlip}\end{split}\end{equation}
where $\varphi_\alpha$ is defined as before, with $\alpha = P$, $1$,... $N$. All the Majorana fermions and $2N$ of the bosons will be gapped out in the strong coupling limit (see Appendix~\ref{app:MajAndVertex}). For the same reasons discussed before this term preserves charge and spin conservation, as well as TRS. 
The remaining gapless bosons are $\phi_{R} = \phi_{\uparrow P} + \sum_{n}\phi_{\uparrow n} $ and $\phi_{L} = \phi_{\downarrow P} + \sum_{n}\phi_{\downarrow n} $, and are described by the minimal edge theory, Eq.~\ref{eq:effectiveLag}, with $\sigma_{sH} = N+\frac{1}{2}$.

\textit{Non-Abelian FTIs.}--- We now ask if Eq.~\ref{eq:effectiveLag} remains stable if $s^z$-conservation is broken, but TRS is preserved.
If the edge is stable, the system with broken $s^z$-conservation is an FTI. 
To determine whether this is the case, we can attempt to gap the bosons in Eq.~\ref{eq:effectiveLag} by condensing the $\varphi$ boson. For the half-integer FQSHs, $\varphi$ can be condensed by a local charge conserving and time-reversal symmetric term $\propto \cos(4\varphi)$. For an FTI,  condensing $\varphi$ must break TRS spontaneously (i.e., there must be a local time-reversal odd operator with an expectation value), since the gapless modes of an FTI are protected against any TRS perturbation. For the half-integer Abelian FQSH, the only local operators that acquire expectation values are of the form, $\cos(4 n (\varphi + \varphi_0))$ where $n\in \mathbb{Z}$, and $\varphi_0$ is an arbitrary constant offset. All these operators are time-reversal even. Thus, the Abelian half-integer FQSHs are not FTIs, in agreement with the results of Ref.~\cite{levin2009fractional}. However, in the non-Abelian case, the operator $i \psi_{\uparrow m} \psi_{\downarrow m} \cos(2 (\varphi + \varphi_0))$ is a local time-reversal odd operator and acquires an expectation value when $\varphi$ is condensed. Eq.~\ref{eq:effectiveLag} is therefore perturbatively stable to breaking $s^z$-conservation when time-reversal is preserved, provided that the bulk state is non-Abelian. The state with broken $s^z$-conservation is a non-Abelian FTI. The minimal edge theory for the non-Abelian FTI is Eq.~\ref{eq:effectiveMagLag}. We now construct this state by adding $s^z$-non-conserving terms to the decoupled edge of a gPf FQSH. 

For an $M = 1$ gPF FQSH, the bosonic charge degrees of freedom of the gPf FTI can be gapped out with the following TRS invariant double electron tunneling term, 
\begin{equation}\begin{split}
\frac{u}{2} \chi^\dagger_{\uparrow 1} \chi^\dagger_{\uparrow 1} \chi_{\downarrow 1}\chi_{ \downarrow 1} = u\cos(4\varphi_{P}),
\label{eq:TRSSInt1}\end{split}\end{equation}
which condenses the $\varphi_{P}$ boson at strong coupling. Unlike before, the Majoranas remain gapless here, so condensing $\varphi$ does not break TRS. Indeed, the only local operators with expectation values are of the form $\cos(4 \varphi_P)$, which are time-reversal even. If we were to additionally gap out the Majorana fermions, time reversal symmetry would be broken, as the local operator $\chi^\dagger_{\uparrow 1} \chi_{\downarrow 1} = i\psi_{\uparrow 1} \psi_{\downarrow 1}e^{i 2 \varphi_{P}}$ would have a non-vanishing expectation value.

For $M > 1$, we have the following time-reversal symmetric combination of spin flips,
\begin{equation}\begin{split}
&\frac{t}{2}\sum_{mm'}[\chi^\dagger_{\uparrow m} \chi_{\downarrow m'} - \chi^\dagger_{\downarrow m} \chi_{\uparrow m'}] + h.c. \\ &= t \sum_{mm'}(i \psi_{\uparrow m} \psi_{\downarrow m'} - i \psi_{\uparrow m'}\psi_{\downarrow m} )\cos(2\varphi_P).
\label{eq:TRSSpinFlip}\end{split}\end{equation}
In the strong coupling limit, this interaction condenses the boson, and induces an effective Majorana mass matrix, $i\bm{\psi}_\uparrow^T \bm{\mathcal{M}}\bm{\psi}_\uparrow$, where $\bm{\psi}_\uparrow = (\psi_{\uparrow 1},..., \psi_{\uparrow m})$ and $\bm{\psi}_\downarrow = (\psi_{\downarrow 1},..., \psi_{\downarrow m})$, and $\bm{\mathcal{M}} = - \bm{\mathcal{M}}^T$ is a $M\times M$ skew matrix. Since the spin-up Majoranas all propagate in the opposite direction of the spin-down Majoranas,
each non-zero eigenvalue of $\bm{\mathcal{M}}$ gaps out a pair of Majorana. An $M \times M$ skew matrix must have at least one zero eigenvalue when $M$ is odd, and a corresponding pair of gapless Majoranas. The fact that this edge preserves TRS is again confirmed by the fact that there are no local time-reversal odd operators with expectation values. In contrast, when $M$ is even and the state is Abelian, there is no guaranteed zero mass, and the edge may be fully gapped.

We can also consider creating a non-Abelian FTI by breaking $s^z$-conservation at the edge of an integer+gPf FQSH with $N$ integer modes and $M$ Majorana modes. As we show in Appendix~\ref{app:IntegerToMajorana} such a system is actually equivalent to a gPf FQSH with $M + 2N$ Majoranas.

\textit{Symmetry Broken Gapped Edges.}--- The gPF FQSHs/FTIs edges can only be gapped out by breaking a symmetry. This can involve, for example, breaking U$(1)$ symmetry, with superconducting (SC) proximity effect, $\chi^\dagger_{\uparrow m} \chi^\dagger_{\downarrow m} = i \psi_{\uparrow m} \psi_{\downarrow m} \cos(2 \theta_P)$, or by breaking U$(1)_\text{s}$ and TRS with an inplane magnetic field, $\chi^\dagger_{\uparrow m} \chi_{\downarrow m} =   i \psi_{\uparrow m} \psi_{\downarrow m} \cos(2 \varphi_P)$. For the SC edge, $\langle i \psi_{\uparrow m} \psi_{\downarrow m} \rangle, \langle \cos(2 \theta_P) \rangle \neq 0$, while for the magnetic edge $\langle i \psi_{\uparrow m} \psi_{\downarrow m} \rangle, \langle \cos(2 \varphi_P) \rangle \neq 0$ (see Appendix~\ref{app:MajAndVertex}).  

An interface between SC and magnetic edges binds a $\mathbb{Z}_4$ parafermion and a Majorana zero mode. The $\mathbb{Z}_4$ parafermion is associated with the bosonic sector. For a domain at $x_{\text{DW}}$ between a SC  ($x < x_{\text{DW}}$) and magnetic ($x > x_{\text{DW}}$) region, the operator $\gamma(x_{\text{DW}}) = e^{i \frac{1}{2}\theta_P (x_{\text{DW}} - \delta)}  e^{i \frac{1}{2}\varphi_P (x_{\text{DW}} + \delta)}$ commutes with the Hamiltonian, where $\delta \rightarrow 0^+$ is small distance\cite{clarke2013exotic, alicea2016topological}. If there are multiple such domain walls, these operators satisfy the $\mathbb{Z}_4$ parafermion algebra,  $\gamma(x_{\text{DW}}) \gamma(x_{\text{DW}}') = \gamma(x_{\text{DW}}') \gamma(x_{\text{DW}}) e^{i \frac{\pi}{2} \text{sgn}(x_{\text{DW}} -x_{\text{DW}}' )} $. The Majorana zero mode is associated with a change in the value of $\langle i \psi_{\uparrow m} \psi_{\downarrow m} \rangle$ across the domain wall\cite{jackiw1976solitons}. Any integer QSH/TI edge modes will also bind an additional Majorana zero modes at a SC and magnetic interface\cite{fu2008superconducting}. 

For a FQSH, a magnetic domain wall where the angle of the inplane magnetic field changes by $\varphi_{\text{B}}$ will bind a charge of $\sigma_{sH} \frac{\varphi_{\text{B}}}{2\pi}$. This can be confirmed by direct calculation. Alternatively, we can ``unfold'' along the edge, turning the FQSH into a single sheet of spin-up FQH with $\sigma_H = \sigma_{sH}$. The unfolding turns the magnetic domain wall into a bulk $\varphi_B$ magnetic flux, which binds charge $\sigma_{sH} \frac{\varphi_{\text{B}}}{2\pi}$ due to the bulk Hall effect. A case worth remarking upon is a domain wall that flips the sign of the magnetic field, $\varphi_{\text{B}} = \pi$, which binds a charge of $1/4$. 

\textit{Relevance to twisted MoTe$_2$}---Ref~\cite{kang2024observation} showed evidence that tMoTe$_2$ has an edge conductance of $\frac{3e^2}{2h}$ at filling $\nu=3$, with no accompanying Hall conductance. This implies that  tMoTe$_2$ is a FQSH with quantized spin Hall conductance of $\sigma_{sH} = \frac{3}{2}$. If true, the spin-resolved Streda formula indicates that the spin Hall conductance also manifests as constant magnetic susceptibility in tMoTe$_2$. In this case, our findings also indicate that if edge interactions are sufficiently strong, tMoTe$_2$ will realize the minimal bosonic edge, with a gap to single electron tunneling. The single electron gap may be detectable via the variation of the Fano factor in shot noise of the tunneling current. Analogous reasoning indicates that strong interactions will also lead to an edge that is gapped to single electrons in the double (triple) quantum spin Hall state that was observed at filling $\nu = 4$ ($6$). If interactions are very weak at the edge of tMoTe$_2$, then spin-up and down edge modes would decouple from each other, and there would be no single electron edge gap. For half-integer FQSH, adding strong Rashba spin-orbit coupling via a proximity effect should make the edge electrically insulating. If tMoTe$_2$ is a non-Abelian FQSH, there will be a remaining pair of Majorana modes that can conduct heat, provided TRS is not spontaneously broken.

It is worth remarking on the possibility that MoTe$_2$ is not fully gapped, but rather has a neutral Fermi surface, as proposed in Refs.~\cite{zhang2024vortex} and~\cite{shi2024excitonic}. Provided there is a bulk charge and spin gap, it is still possible to have well-defined edge modes that carry charge and are described by a Luttinger liquid\cite{shytov1998tunneling, levitov2001effective, barkeshli2015particle}. Due to the gapless neutral fermions in the bulk, this edge will have local fermion operators. The presence of gapless bulk fermions will also lead to different heat transport signatures compared to a system with only gapless edge excitations. 

\textit{Acknowledgements}--- We thank Kin-Fai Mak for his assistance in the inception of this project. We also thank Aidan P Reddy, Tixuan Tan, Ramanjit Sohal, Srinivas Raghu, Steve Kivelson, Hart Goldman, Erez Berg and Netanel Lindner for useful conversations. JMM and TD were supported by a startup fund at Stanford University. AS was supported by grants from the ERC under the European Union’s Horizon 2020 research and innovation programme (Grant Agreements LEGOTOP No. 788715 ), the DFG (CRC/Transregio 183, EI 519/71), and by the ISF Quantum Science and Technology (2074/19).

\bibliography{FQSH_Edge.bib}

\begin{thebibliography}{70}%
\makeatletter
\providecommand \@ifxundefined [1]{%
 \@ifx{#1\undefined}
}%
\providecommand \@ifnum [1]{%
 \ifnum #1\expandafter \@firstoftwo
 \else \expandafter \@secondoftwo
 \fi
}%
\providecommand \@ifx [1]{%
 \ifx #1\expandafter \@firstoftwo
 \else \expandafter \@secondoftwo
 \fi
}%
\providecommand \natexlab [1]{#1}%
\providecommand \enquote  [1]{``#1''}%
\providecommand \bibnamefont  [1]{#1}%
\providecommand \bibfnamefont [1]{#1}%
\providecommand \citenamefont [1]{#1}%
\providecommand \href@noop [0]{\@secondoftwo}%
\providecommand \href [0]{\begingroup \@sanitize@url \@href}%
\providecommand \@href[1]{\@@startlink{#1}\@@href}%
\providecommand \@@href[1]{\endgroup#1\@@endlink}%
\providecommand \@sanitize@url [0]{\catcode `\\12\catcode `\$12\catcode
  `\&12\catcode `\#12\catcode `\^12\catcode `\_12\catcode `\%12\relax}%
\providecommand \@@startlink[1]{}%
\providecommand \@@endlink[0]{}%
\providecommand \url  [0]{\begingroup\@sanitize@url \@url }%
\providecommand \@url [1]{\endgroup\@href {#1}{\urlprefix }}%
\providecommand \urlprefix  [0]{URL }%
\providecommand \Eprint [0]{\href }%
\providecommand \doibase [0]{http://dx.doi.org/}%
\providecommand \selectlanguage [0]{\@gobble}%
\providecommand \bibinfo  [0]{\@secondoftwo}%
\providecommand \bibfield  [0]{\@secondoftwo}%
\providecommand \translation [1]{[#1]}%
\providecommand \BibitemOpen [0]{}%
\providecommand \bibitemStop [0]{}%
\providecommand \bibitemNoStop [0]{.\EOS\space}%
\providecommand \EOS [0]{\spacefactor3000\relax}%
\providecommand \BibitemShut  [1]{\csname bibitem#1\endcsname}%
\let\auto@bib@innerbib\@empty
\bibitem [{\citenamefont {Stormer}\ \emph {et~al.}(1999)\citenamefont
  {Stormer}, \citenamefont {Tsui},\ and\ \citenamefont
  {Gossard}}]{stormer1999fractional}%
  \BibitemOpen
  \bibfield  {author} {\bibinfo {author} {\bibfnamefont {H.~L.}\ \bibnamefont
  {Stormer}}, \bibinfo {author} {\bibfnamefont {D.~C.}\ \bibnamefont {Tsui}}, \
  and\ \bibinfo {author} {\bibfnamefont {A.~C.}\ \bibnamefont {Gossard}},\
  }\href@noop {} {\bibfield  {journal} {\bibinfo  {journal} {Reviews of Modern
  Physics}\ }\textbf {\bibinfo {volume} {71}},\ \bibinfo {pages} {S298}
  (\bibinfo {year} {1999})}\BibitemShut {NoStop}%
\bibitem [{\citenamefont {Stern}(2008)}]{stern2008anyons}%
  \BibitemOpen
  \bibfield  {author} {\bibinfo {author} {\bibfnamefont {A.}~\bibnamefont
  {Stern}},\ }\href@noop {} {\bibfield  {journal} {\bibinfo  {journal} {Annals
  of Physics}\ }\textbf {\bibinfo {volume} {323}},\ \bibinfo {pages} {204}
  (\bibinfo {year} {2008})}\BibitemShut {NoStop}%
\bibitem [{\citenamefont {Cage}\ \emph {et~al.}(2012)\citenamefont {Cage},
  \citenamefont {Klitzing}, \citenamefont {Chang}, \citenamefont {Duncan},
  \citenamefont {Haldane}, \citenamefont {Laughlin}, \citenamefont {Pruisken},\
  and\ \citenamefont {Thouless}}]{cage2012quantum}%
  \BibitemOpen
  \bibfield  {author} {\bibinfo {author} {\bibfnamefont {M.~E.}\ \bibnamefont
  {Cage}}, \bibinfo {author} {\bibfnamefont {K.}~\bibnamefont {Klitzing}},
  \bibinfo {author} {\bibfnamefont {A.}~\bibnamefont {Chang}}, \bibinfo
  {author} {\bibfnamefont {F.}~\bibnamefont {Duncan}}, \bibinfo {author}
  {\bibfnamefont {M.}~\bibnamefont {Haldane}}, \bibinfo {author} {\bibfnamefont
  {R.~B.}\ \bibnamefont {Laughlin}}, \bibinfo {author} {\bibfnamefont
  {A.}~\bibnamefont {Pruisken}}, \ and\ \bibinfo {author} {\bibfnamefont
  {D.}~\bibnamefont {Thouless}},\ }\href@noop {} {\emph {\bibinfo {title} {The
  quantum Hall effect}}}\ (\bibinfo  {publisher} {Springer Science \& Business
  Media},\ \bibinfo {year} {2012})\BibitemShut {NoStop}%
\bibitem [{\citenamefont {Sarma}\ \emph {et~al.}(2005)\citenamefont {Sarma},
  \citenamefont {Freedman},\ and\ \citenamefont
  {Nayak}}]{sarma2005topologically}%
  \BibitemOpen
  \bibfield  {author} {\bibinfo {author} {\bibfnamefont {S.~D.}\ \bibnamefont
  {Sarma}}, \bibinfo {author} {\bibfnamefont {M.}~\bibnamefont {Freedman}}, \
  and\ \bibinfo {author} {\bibfnamefont {C.}~\bibnamefont {Nayak}},\
  }\href@noop {} {\bibfield  {journal} {\bibinfo  {journal} {Physical review
  letters}\ }\textbf {\bibinfo {volume} {94}},\ \bibinfo {pages} {166802}
  (\bibinfo {year} {2005})}\BibitemShut {NoStop}%
\bibitem [{\citenamefont {Nayak}\ \emph {et~al.}(2008)\citenamefont {Nayak},
  \citenamefont {Simon}, \citenamefont {Stern}, \citenamefont {Freedman},\ and\
  \citenamefont {Sarma}}]{nayak2008non}%
  \BibitemOpen
  \bibfield  {author} {\bibinfo {author} {\bibfnamefont {C.}~\bibnamefont
  {Nayak}}, \bibinfo {author} {\bibfnamefont {S.~H.}\ \bibnamefont {Simon}},
  \bibinfo {author} {\bibfnamefont {A.}~\bibnamefont {Stern}}, \bibinfo
  {author} {\bibfnamefont {M.}~\bibnamefont {Freedman}}, \ and\ \bibinfo
  {author} {\bibfnamefont {S.~D.}\ \bibnamefont {Sarma}},\ }\href@noop {}
  {\bibfield  {journal} {\bibinfo  {journal} {Reviews of Modern Physics}\
  }\textbf {\bibinfo {volume} {80}},\ \bibinfo {pages} {1083} (\bibinfo {year}
  {2008})}\BibitemShut {NoStop}%
\bibitem [{\citenamefont {Andrei}\ and\ \citenamefont
  {MacDonald}(2020)}]{andrei2020graphene}%
  \BibitemOpen
  \bibfield  {author} {\bibinfo {author} {\bibfnamefont {E.~Y.}\ \bibnamefont
  {Andrei}}\ and\ \bibinfo {author} {\bibfnamefont {A.~H.}\ \bibnamefont
  {MacDonald}},\ }\href@noop {} {\bibfield  {journal} {\bibinfo  {journal}
  {Nature materials}\ }\textbf {\bibinfo {volume} {19}},\ \bibinfo {pages}
  {1265} (\bibinfo {year} {2020})}\BibitemShut {NoStop}%
\bibitem [{\citenamefont {Balents}\ \emph {et~al.}(2020)\citenamefont
  {Balents}, \citenamefont {Dean}, \citenamefont {Efetov},\ and\ \citenamefont
  {Young}}]{balents2020superconductivity}%
  \BibitemOpen
  \bibfield  {author} {\bibinfo {author} {\bibfnamefont {L.}~\bibnamefont
  {Balents}}, \bibinfo {author} {\bibfnamefont {C.~R.}\ \bibnamefont {Dean}},
  \bibinfo {author} {\bibfnamefont {D.~K.}\ \bibnamefont {Efetov}}, \ and\
  \bibinfo {author} {\bibfnamefont {A.~F.}\ \bibnamefont {Young}},\ }\href@noop
  {} {\bibfield  {journal} {\bibinfo  {journal} {Nature Physics}\ }\textbf
  {\bibinfo {volume} {16}},\ \bibinfo {pages} {725} (\bibinfo {year}
  {2020})}\BibitemShut {NoStop}%
\bibitem [{\citenamefont {He}\ \emph {et~al.}(2021)\citenamefont {He},
  \citenamefont {Zhou}, \citenamefont {Ye}, \citenamefont {Cho}, \citenamefont
  {Jeong}, \citenamefont {Meng},\ and\ \citenamefont {Wang}}]{he2021moire}%
  \BibitemOpen
  \bibfield  {author} {\bibinfo {author} {\bibfnamefont {F.}~\bibnamefont
  {He}}, \bibinfo {author} {\bibfnamefont {Y.}~\bibnamefont {Zhou}}, \bibinfo
  {author} {\bibfnamefont {Z.}~\bibnamefont {Ye}}, \bibinfo {author}
  {\bibfnamefont {S.-H.}\ \bibnamefont {Cho}}, \bibinfo {author} {\bibfnamefont
  {J.}~\bibnamefont {Jeong}}, \bibinfo {author} {\bibfnamefont
  {X.}~\bibnamefont {Meng}}, \ and\ \bibinfo {author} {\bibfnamefont
  {Y.}~\bibnamefont {Wang}},\ }\href@noop {} {\bibfield  {journal} {\bibinfo
  {journal} {ACS nano}\ }\textbf {\bibinfo {volume} {15}},\ \bibinfo {pages}
  {5944} (\bibinfo {year} {2021})}\BibitemShut {NoStop}%
\bibitem [{\citenamefont {Mak}\ and\ \citenamefont
  {Shan}(2022)}]{mak2022semiconductor}%
  \BibitemOpen
  \bibfield  {author} {\bibinfo {author} {\bibfnamefont {K.~F.}\ \bibnamefont
  {Mak}}\ and\ \bibinfo {author} {\bibfnamefont {J.}~\bibnamefont {Shan}},\
  }\href@noop {} {\bibfield  {journal} {\bibinfo  {journal} {Nature
  Nanotechnology}\ }\textbf {\bibinfo {volume} {17}},\ \bibinfo {pages} {686}
  (\bibinfo {year} {2022})}\BibitemShut {NoStop}%
\bibitem [{\citenamefont {Regnault}\ and\ \citenamefont
  {Bernevig}(2011)}]{regnault2011fractional}%
  \BibitemOpen
  \bibfield  {author} {\bibinfo {author} {\bibfnamefont {N.}~\bibnamefont
  {Regnault}}\ and\ \bibinfo {author} {\bibfnamefont {B.~A.}\ \bibnamefont
  {Bernevig}},\ }\href@noop {} {\bibfield  {journal} {\bibinfo  {journal}
  {Physical Review X}\ }\textbf {\bibinfo {volume} {1}},\ \bibinfo {pages}
  {021014} (\bibinfo {year} {2011})}\BibitemShut {NoStop}%
\bibitem [{\citenamefont {Neupert}\ \emph {et~al.}(2011)\citenamefont
  {Neupert}, \citenamefont {Santos}, \citenamefont {Chamon},\ and\
  \citenamefont {Mudry}}]{neupert2011fractional}%
  \BibitemOpen
  \bibfield  {author} {\bibinfo {author} {\bibfnamefont {T.}~\bibnamefont
  {Neupert}}, \bibinfo {author} {\bibfnamefont {L.}~\bibnamefont {Santos}},
  \bibinfo {author} {\bibfnamefont {C.}~\bibnamefont {Chamon}}, \ and\ \bibinfo
  {author} {\bibfnamefont {C.}~\bibnamefont {Mudry}},\ }\href@noop {}
  {\bibfield  {journal} {\bibinfo  {journal} {Physical review letters}\
  }\textbf {\bibinfo {volume} {106}},\ \bibinfo {pages} {236804} (\bibinfo
  {year} {2011})}\BibitemShut {NoStop}%
\bibitem [{\citenamefont {Tang}\ \emph {et~al.}(2011)\citenamefont {Tang},
  \citenamefont {Mei},\ and\ \citenamefont {Wen}}]{tang2011high}%
  \BibitemOpen
  \bibfield  {author} {\bibinfo {author} {\bibfnamefont {E.}~\bibnamefont
  {Tang}}, \bibinfo {author} {\bibfnamefont {J.-W.}\ \bibnamefont {Mei}}, \
  and\ \bibinfo {author} {\bibfnamefont {X.-G.}\ \bibnamefont {Wen}},\
  }\href@noop {} {\bibfield  {journal} {\bibinfo  {journal} {Physical review
  letters}\ }\textbf {\bibinfo {volume} {106}},\ \bibinfo {pages} {236802}
  (\bibinfo {year} {2011})}\BibitemShut {NoStop}%
\bibitem [{\citenamefont {Sheng}\ \emph {et~al.}(2011)\citenamefont {Sheng},
  \citenamefont {Gu}, \citenamefont {Sun},\ and\ \citenamefont
  {Sheng}}]{sheng2011fractional}%
  \BibitemOpen
  \bibfield  {author} {\bibinfo {author} {\bibfnamefont {D.}~\bibnamefont
  {Sheng}}, \bibinfo {author} {\bibfnamefont {Z.-C.}\ \bibnamefont {Gu}},
  \bibinfo {author} {\bibfnamefont {K.}~\bibnamefont {Sun}}, \ and\ \bibinfo
  {author} {\bibfnamefont {L.}~\bibnamefont {Sheng}},\ }\href@noop {}
  {\bibfield  {journal} {\bibinfo  {journal} {Nature communications}\ }\textbf
  {\bibinfo {volume} {2}},\ \bibinfo {pages} {389} (\bibinfo {year}
  {2011})}\BibitemShut {NoStop}%
\bibitem [{\citenamefont {Wu}\ \emph {et~al.}(2012)\citenamefont {Wu},
  \citenamefont {Bernevig},\ and\ \citenamefont {Regnault}}]{wu2012zoology}%
  \BibitemOpen
  \bibfield  {author} {\bibinfo {author} {\bibfnamefont {Y.-L.}\ \bibnamefont
  {Wu}}, \bibinfo {author} {\bibfnamefont {B.~A.}\ \bibnamefont {Bernevig}}, \
  and\ \bibinfo {author} {\bibfnamefont {N.}~\bibnamefont {Regnault}},\
  }\href@noop {} {\bibfield  {journal} {\bibinfo  {journal} {Physical Review
  B}\ }\textbf {\bibinfo {volume} {85}},\ \bibinfo {pages} {075116} (\bibinfo
  {year} {2012})}\BibitemShut {NoStop}%
\bibitem [{\citenamefont {Cai}\ \emph {et~al.}(2023)\citenamefont {Cai},
  \citenamefont {Anderson}, \citenamefont {Wang}, \citenamefont {Zhang},
  \citenamefont {Liu}, \citenamefont {Holtzmann}, \citenamefont {Zhang},
  \citenamefont {Fan}, \citenamefont {Taniguchi}, \citenamefont {Watanabe}
  \emph {et~al.}}]{cai2023signatures}%
  \BibitemOpen
  \bibfield  {author} {\bibinfo {author} {\bibfnamefont {J.}~\bibnamefont
  {Cai}}, \bibinfo {author} {\bibfnamefont {E.}~\bibnamefont {Anderson}},
  \bibinfo {author} {\bibfnamefont {C.}~\bibnamefont {Wang}}, \bibinfo {author}
  {\bibfnamefont {X.}~\bibnamefont {Zhang}}, \bibinfo {author} {\bibfnamefont
  {X.}~\bibnamefont {Liu}}, \bibinfo {author} {\bibfnamefont {W.}~\bibnamefont
  {Holtzmann}}, \bibinfo {author} {\bibfnamefont {Y.}~\bibnamefont {Zhang}},
  \bibinfo {author} {\bibfnamefont {F.}~\bibnamefont {Fan}}, \bibinfo {author}
  {\bibfnamefont {T.}~\bibnamefont {Taniguchi}}, \bibinfo {author}
  {\bibfnamefont {K.}~\bibnamefont {Watanabe}},  \emph {et~al.},\ }\href@noop
  {} {\bibfield  {journal} {\bibinfo  {journal} {Nature}\ }\textbf {\bibinfo
  {volume} {622}},\ \bibinfo {pages} {63} (\bibinfo {year} {2023})}\BibitemShut
  {NoStop}%
\bibitem [{\citenamefont {Park}\ \emph {et~al.}(2023)\citenamefont {Park},
  \citenamefont {Cai}, \citenamefont {Anderson}, \citenamefont {Zhang},
  \citenamefont {Zhu}, \citenamefont {Liu}, \citenamefont {Wang}, \citenamefont
  {Holtzmann}, \citenamefont {Hu}, \citenamefont {Liu} \emph
  {et~al.}}]{park2023observation}%
  \BibitemOpen
  \bibfield  {author} {\bibinfo {author} {\bibfnamefont {H.}~\bibnamefont
  {Park}}, \bibinfo {author} {\bibfnamefont {J.}~\bibnamefont {Cai}}, \bibinfo
  {author} {\bibfnamefont {E.}~\bibnamefont {Anderson}}, \bibinfo {author}
  {\bibfnamefont {Y.}~\bibnamefont {Zhang}}, \bibinfo {author} {\bibfnamefont
  {J.}~\bibnamefont {Zhu}}, \bibinfo {author} {\bibfnamefont {X.}~\bibnamefont
  {Liu}}, \bibinfo {author} {\bibfnamefont {C.}~\bibnamefont {Wang}}, \bibinfo
  {author} {\bibfnamefont {W.}~\bibnamefont {Holtzmann}}, \bibinfo {author}
  {\bibfnamefont {C.}~\bibnamefont {Hu}}, \bibinfo {author} {\bibfnamefont
  {Z.}~\bibnamefont {Liu}},  \emph {et~al.},\ }\href@noop {} {\bibfield
  {journal} {\bibinfo  {journal} {Nature}\ }\textbf {\bibinfo {volume} {622}},\
  \bibinfo {pages} {74} (\bibinfo {year} {2023})}\BibitemShut {NoStop}%
\bibitem [{\citenamefont {Zeng}\ \emph {et~al.}(2023)\citenamefont {Zeng},
  \citenamefont {Xia}, \citenamefont {Kang}, \citenamefont {Zhu}, \citenamefont
  {Kn{\"u}ppel}, \citenamefont {Vaswani}, \citenamefont {Watanabe},
  \citenamefont {Taniguchi}, \citenamefont {Mak},\ and\ \citenamefont
  {Shan}}]{zeng2023thermodynamic}%
  \BibitemOpen
  \bibfield  {author} {\bibinfo {author} {\bibfnamefont {Y.}~\bibnamefont
  {Zeng}}, \bibinfo {author} {\bibfnamefont {Z.}~\bibnamefont {Xia}}, \bibinfo
  {author} {\bibfnamefont {K.}~\bibnamefont {Kang}}, \bibinfo {author}
  {\bibfnamefont {J.}~\bibnamefont {Zhu}}, \bibinfo {author} {\bibfnamefont
  {P.}~\bibnamefont {Kn{\"u}ppel}}, \bibinfo {author} {\bibfnamefont
  {C.}~\bibnamefont {Vaswani}}, \bibinfo {author} {\bibfnamefont
  {K.}~\bibnamefont {Watanabe}}, \bibinfo {author} {\bibfnamefont
  {T.}~\bibnamefont {Taniguchi}}, \bibinfo {author} {\bibfnamefont {K.~F.}\
  \bibnamefont {Mak}}, \ and\ \bibinfo {author} {\bibfnamefont
  {J.}~\bibnamefont {Shan}},\ }\href@noop {} {\bibfield  {journal} {\bibinfo
  {journal} {Nature}\ }\textbf {\bibinfo {volume} {622}},\ \bibinfo {pages}
  {69} (\bibinfo {year} {2023})}\BibitemShut {NoStop}%
\bibitem [{\citenamefont {Xu}\ \emph {et~al.}(2023)\citenamefont {Xu},
  \citenamefont {Sun}, \citenamefont {Jia}, \citenamefont {Liu}, \citenamefont
  {Xu}, \citenamefont {Li}, \citenamefont {Gu}, \citenamefont {Watanabe},
  \citenamefont {Taniguchi}, \citenamefont {Tong} \emph
  {et~al.}}]{xu2023observation}%
  \BibitemOpen
  \bibfield  {author} {\bibinfo {author} {\bibfnamefont {F.}~\bibnamefont
  {Xu}}, \bibinfo {author} {\bibfnamefont {Z.}~\bibnamefont {Sun}}, \bibinfo
  {author} {\bibfnamefont {T.}~\bibnamefont {Jia}}, \bibinfo {author}
  {\bibfnamefont {C.}~\bibnamefont {Liu}}, \bibinfo {author} {\bibfnamefont
  {C.}~\bibnamefont {Xu}}, \bibinfo {author} {\bibfnamefont {C.}~\bibnamefont
  {Li}}, \bibinfo {author} {\bibfnamefont {Y.}~\bibnamefont {Gu}}, \bibinfo
  {author} {\bibfnamefont {K.}~\bibnamefont {Watanabe}}, \bibinfo {author}
  {\bibfnamefont {T.}~\bibnamefont {Taniguchi}}, \bibinfo {author}
  {\bibfnamefont {B.}~\bibnamefont {Tong}},  \emph {et~al.},\ }\href@noop {}
  {\bibfield  {journal} {\bibinfo  {journal} {Physical Review X}\ }\textbf
  {\bibinfo {volume} {13}},\ \bibinfo {pages} {031037} (\bibinfo {year}
  {2023})}\BibitemShut {NoStop}%
\bibitem [{\citenamefont {Lu}\ \emph {et~al.}(2024)\citenamefont {Lu},
  \citenamefont {Han}, \citenamefont {Yao}, \citenamefont {Reddy},
  \citenamefont {Yang}, \citenamefont {Seo}, \citenamefont {Watanabe},
  \citenamefont {Taniguchi}, \citenamefont {Fu},\ and\ \citenamefont
  {Ju}}]{lu2024fractional}%
  \BibitemOpen
  \bibfield  {author} {\bibinfo {author} {\bibfnamefont {Z.}~\bibnamefont
  {Lu}}, \bibinfo {author} {\bibfnamefont {T.}~\bibnamefont {Han}}, \bibinfo
  {author} {\bibfnamefont {Y.}~\bibnamefont {Yao}}, \bibinfo {author}
  {\bibfnamefont {A.~P.}\ \bibnamefont {Reddy}}, \bibinfo {author}
  {\bibfnamefont {J.}~\bibnamefont {Yang}}, \bibinfo {author} {\bibfnamefont
  {J.}~\bibnamefont {Seo}}, \bibinfo {author} {\bibfnamefont {K.}~\bibnamefont
  {Watanabe}}, \bibinfo {author} {\bibfnamefont {T.}~\bibnamefont {Taniguchi}},
  \bibinfo {author} {\bibfnamefont {L.}~\bibnamefont {Fu}}, \ and\ \bibinfo
  {author} {\bibfnamefont {L.}~\bibnamefont {Ju}},\ }\href@noop {} {\bibfield
  {journal} {\bibinfo  {journal} {Nature}\ }\textbf {\bibinfo {volume} {626}},\
  \bibinfo {pages} {759} (\bibinfo {year} {2024})}\BibitemShut {NoStop}%
\bibitem [{\citenamefont {Stern}(2016)}]{FTI_review}%
  \BibitemOpen
  \bibfield  {author} {\bibinfo {author} {\bibfnamefont {A.}~\bibnamefont
  {Stern}},\ }\href {\doibase 10.1146/annurev-conmatphys-031115-011559}
  {\bibfield  {journal} {\bibinfo  {journal} {Annual Review of Condensed Matter
  Physics}\ }\textbf {\bibinfo {volume} {7}},\ \bibinfo {pages} {349} (\bibinfo
  {year} {2016})},\ \Eprint
  {http://arxiv.org/abs/https://doi.org/10.1146/annurev-conmatphys-031115-011559}
  {https://doi.org/10.1146/annurev-conmatphys-031115-011559} \BibitemShut
  {NoStop}%
\bibitem [{\citenamefont {Kang}\ \emph {et~al.}(2024)\citenamefont {Kang},
  \citenamefont {Shen}, \citenamefont {Qiu}, \citenamefont {Watanabe},
  \citenamefont {Taniguchi}, \citenamefont {Shan},\ and\ \citenamefont
  {Mak}}]{kang2024observation}%
  \BibitemOpen
  \bibfield  {author} {\bibinfo {author} {\bibfnamefont {K.}~\bibnamefont
  {Kang}}, \bibinfo {author} {\bibfnamefont {B.}~\bibnamefont {Shen}}, \bibinfo
  {author} {\bibfnamefont {Y.}~\bibnamefont {Qiu}}, \bibinfo {author}
  {\bibfnamefont {K.}~\bibnamefont {Watanabe}}, \bibinfo {author}
  {\bibfnamefont {T.}~\bibnamefont {Taniguchi}}, \bibinfo {author}
  {\bibfnamefont {J.}~\bibnamefont {Shan}}, \ and\ \bibinfo {author}
  {\bibfnamefont {K.~F.}\ \bibnamefont {Mak}},\ }\href@noop {} {\bibfield
  {journal} {\bibinfo  {journal} {arXiv preprint arXiv:2402.03294}\ } (\bibinfo
  {year} {2024})}\BibitemShut {NoStop}%
\bibitem [{\citenamefont {Levin}\ and\ \citenamefont
  {Stern}(2009)}]{levin2009fractional}%
  \BibitemOpen
  \bibfield  {author} {\bibinfo {author} {\bibfnamefont {M.}~\bibnamefont
  {Levin}}\ and\ \bibinfo {author} {\bibfnamefont {A.}~\bibnamefont {Stern}},\
  }\href@noop {} {\bibfield  {journal} {\bibinfo  {journal} {Physical review
  letters}\ }\textbf {\bibinfo {volume} {103}},\ \bibinfo {pages} {196803}
  (\bibinfo {year} {2009})}\BibitemShut {NoStop}%
\bibitem [{\citenamefont {Xiao}\ \emph {et~al.}(2012)\citenamefont {Xiao},
  \citenamefont {Liu}, \citenamefont {Feng}, \citenamefont {Xu},\ and\
  \citenamefont {Yao}}]{xiao2012coupled}%
  \BibitemOpen
  \bibfield  {author} {\bibinfo {author} {\bibfnamefont {D.}~\bibnamefont
  {Xiao}}, \bibinfo {author} {\bibfnamefont {G.-B.}\ \bibnamefont {Liu}},
  \bibinfo {author} {\bibfnamefont {W.}~\bibnamefont {Feng}}, \bibinfo {author}
  {\bibfnamefont {X.}~\bibnamefont {Xu}}, \ and\ \bibinfo {author}
  {\bibfnamefont {W.}~\bibnamefont {Yao}},\ }\href@noop {} {\bibfield
  {journal} {\bibinfo  {journal} {Physical review letters}\ }\textbf {\bibinfo
  {volume} {108}},\ \bibinfo {pages} {196802} (\bibinfo {year}
  {2012})}\BibitemShut {NoStop}%
\bibitem [{\citenamefont {Wu}\ \emph {et~al.}(2019)\citenamefont {Wu},
  \citenamefont {Lovorn}, \citenamefont {Tutuc}, \citenamefont {Martin},\ and\
  \citenamefont {MacDonald}}]{wu2019topological}%
  \BibitemOpen
  \bibfield  {author} {\bibinfo {author} {\bibfnamefont {F.}~\bibnamefont
  {Wu}}, \bibinfo {author} {\bibfnamefont {T.}~\bibnamefont {Lovorn}}, \bibinfo
  {author} {\bibfnamefont {E.}~\bibnamefont {Tutuc}}, \bibinfo {author}
  {\bibfnamefont {I.}~\bibnamefont {Martin}}, \ and\ \bibinfo {author}
  {\bibfnamefont {A.}~\bibnamefont {MacDonald}},\ }\href@noop {} {\bibfield
  {journal} {\bibinfo  {journal} {Physical review letters}\ }\textbf {\bibinfo
  {volume} {122}},\ \bibinfo {pages} {086402} (\bibinfo {year}
  {2019})}\BibitemShut {NoStop}%
\bibitem [{\citenamefont {Devakul}\ \emph {et~al.}(2021)\citenamefont
  {Devakul}, \citenamefont {Cr{\'e}pel}, \citenamefont {Zhang},\ and\
  \citenamefont {Fu}}]{devakul2021magic}%
  \BibitemOpen
  \bibfield  {author} {\bibinfo {author} {\bibfnamefont {T.}~\bibnamefont
  {Devakul}}, \bibinfo {author} {\bibfnamefont {V.}~\bibnamefont {Cr{\'e}pel}},
  \bibinfo {author} {\bibfnamefont {Y.}~\bibnamefont {Zhang}}, \ and\ \bibinfo
  {author} {\bibfnamefont {L.}~\bibnamefont {Fu}},\ }\href@noop {} {\bibfield
  {journal} {\bibinfo  {journal} {Nature communications}\ }\textbf {\bibinfo
  {volume} {12}},\ \bibinfo {pages} {6730} (\bibinfo {year}
  {2021})}\BibitemShut {NoStop}%
\bibitem [{\citenamefont {Bernevig}\ and\ \citenamefont
  {Zhang}(2006)}]{bernevig2006quantum}%
  \BibitemOpen
  \bibfield  {author} {\bibinfo {author} {\bibfnamefont {B.~A.}\ \bibnamefont
  {Bernevig}}\ and\ \bibinfo {author} {\bibfnamefont {S.-C.}\ \bibnamefont
  {Zhang}},\ }\href@noop {} {\bibfield  {journal} {\bibinfo  {journal}
  {Physical review letters}\ }\textbf {\bibinfo {volume} {96}},\ \bibinfo
  {pages} {106802} (\bibinfo {year} {2006})}\BibitemShut {NoStop}%
\bibitem [{\citenamefont {Qi}(2011)}]{qi2011generic}%
  \BibitemOpen
  \bibfield  {author} {\bibinfo {author} {\bibfnamefont {X.-L.}\ \bibnamefont
  {Qi}},\ }\href@noop {} {\bibfield  {journal} {\bibinfo  {journal} {Physical
  review letters}\ }\textbf {\bibinfo {volume} {107}},\ \bibinfo {pages}
  {126803} (\bibinfo {year} {2011})}\BibitemShut {NoStop}%
\bibitem [{\citenamefont {Levin}\ \emph {et~al.}(2011)\citenamefont {Levin},
  \citenamefont {Burnell}, \citenamefont {Koch-Janusz},\ and\ \citenamefont
  {Stern}}]{levin2011exactly}%
  \BibitemOpen
  \bibfield  {author} {\bibinfo {author} {\bibfnamefont {M.}~\bibnamefont
  {Levin}}, \bibinfo {author} {\bibfnamefont {F.}~\bibnamefont {Burnell}},
  \bibinfo {author} {\bibfnamefont {M.}~\bibnamefont {Koch-Janusz}}, \ and\
  \bibinfo {author} {\bibfnamefont {A.}~\bibnamefont {Stern}},\ }\href@noop {}
  {\bibfield  {journal} {\bibinfo  {journal} {Physical Review B}\ }\textbf
  {\bibinfo {volume} {84}},\ \bibinfo {pages} {235145} (\bibinfo {year}
  {2011})}\BibitemShut {NoStop}%
\bibitem [{\citenamefont {Lu}\ and\ \citenamefont
  {Ran}(2012)}]{lu2012symmetry}%
  \BibitemOpen
  \bibfield  {author} {\bibinfo {author} {\bibfnamefont {Y.-M.}\ \bibnamefont
  {Lu}}\ and\ \bibinfo {author} {\bibfnamefont {Y.}~\bibnamefont {Ran}},\
  }\href@noop {} {\bibfield  {journal} {\bibinfo  {journal} {Physical Review
  B}\ }\textbf {\bibinfo {volume} {85}},\ \bibinfo {pages} {165134} (\bibinfo
  {year} {2012})}\BibitemShut {NoStop}%
\bibitem [{\citenamefont {Levin}\ and\ \citenamefont
  {Stern}(2012)}]{Levin_2012}%
  \BibitemOpen
  \bibfield  {author} {\bibinfo {author} {\bibfnamefont {M.}~\bibnamefont
  {Levin}}\ and\ \bibinfo {author} {\bibfnamefont {A.}~\bibnamefont {Stern}},\
  }\href {\doibase 10.1103/physrevb.86.115131} {\bibfield  {journal} {\bibinfo
  {journal} {Physical Review B}\ }\textbf {\bibinfo {volume} {86}} (\bibinfo
  {year} {2012}),\ 10.1103/physrevb.86.115131}\BibitemShut {NoStop}%
\bibitem [{\citenamefont {Neupert}\ \emph {et~al.}(2015)\citenamefont
  {Neupert}, \citenamefont {Chamon}, \citenamefont {Iadecola}, \citenamefont
  {Santos},\ and\ \citenamefont {Mudry}}]{neupert2015fractional}%
  \BibitemOpen
  \bibfield  {author} {\bibinfo {author} {\bibfnamefont {T.}~\bibnamefont
  {Neupert}}, \bibinfo {author} {\bibfnamefont {C.}~\bibnamefont {Chamon}},
  \bibinfo {author} {\bibfnamefont {T.}~\bibnamefont {Iadecola}}, \bibinfo
  {author} {\bibfnamefont {L.~H.}\ \bibnamefont {Santos}}, \ and\ \bibinfo
  {author} {\bibfnamefont {C.}~\bibnamefont {Mudry}},\ }\href@noop {}
  {\bibfield  {journal} {\bibinfo  {journal} {Physica Scripta}\ }\textbf
  {\bibinfo {volume} {2015}},\ \bibinfo {pages} {014005} (\bibinfo {year}
  {2015})}\BibitemShut {NoStop}%
\bibitem [{\citenamefont {Cappelli}\ and\ \citenamefont
  {Randellini}(2013)}]{cappelli2013partition}%
  \BibitemOpen
  \bibfield  {author} {\bibinfo {author} {\bibfnamefont {A.}~\bibnamefont
  {Cappelli}}\ and\ \bibinfo {author} {\bibfnamefont {E.}~\bibnamefont
  {Randellini}},\ }\href@noop {} {\bibfield  {journal} {\bibinfo  {journal}
  {Journal of High Energy Physics}\ }\textbf {\bibinfo {volume} {2013}},\
  \bibinfo {pages} {1} (\bibinfo {year} {2013})}\BibitemShut {NoStop}%
\bibitem [{\citenamefont {Cappelli}\ and\ \citenamefont
  {Randellini}(2015)}]{cappelli2015stability}%
  \BibitemOpen
  \bibfield  {author} {\bibinfo {author} {\bibfnamefont {A.}~\bibnamefont
  {Cappelli}}\ and\ \bibinfo {author} {\bibfnamefont {E.}~\bibnamefont
  {Randellini}},\ }\href@noop {} {\bibfield  {journal} {\bibinfo  {journal}
  {Journal of Physics A: Mathematical and Theoretical}\ }\textbf {\bibinfo
  {volume} {48}},\ \bibinfo {pages} {105404} (\bibinfo {year}
  {2015})}\BibitemShut {NoStop}%
\bibitem [{\citenamefont {Willett}\ \emph {et~al.}(1987)\citenamefont
  {Willett}, \citenamefont {Eisenstein}, \citenamefont {St{\"o}rmer},
  \citenamefont {Tsui}, \citenamefont {Gossard},\ and\ \citenamefont
  {English}}]{willett1987observation}%
  \BibitemOpen
  \bibfield  {author} {\bibinfo {author} {\bibfnamefont {R.}~\bibnamefont
  {Willett}}, \bibinfo {author} {\bibfnamefont {J.~P.}\ \bibnamefont
  {Eisenstein}}, \bibinfo {author} {\bibfnamefont {H.~L.}\ \bibnamefont
  {St{\"o}rmer}}, \bibinfo {author} {\bibfnamefont {D.~C.}\ \bibnamefont
  {Tsui}}, \bibinfo {author} {\bibfnamefont {A.~C.}\ \bibnamefont {Gossard}}, \
  and\ \bibinfo {author} {\bibfnamefont {J.}~\bibnamefont {English}},\
  }\href@noop {} {\bibfield  {journal} {\bibinfo  {journal} {Physical review
  letters}\ }\textbf {\bibinfo {volume} {59}},\ \bibinfo {pages} {1776}
  (\bibinfo {year} {1987})}\BibitemShut {NoStop}%
\bibitem [{\citenamefont {Willett}(2013)}]{willett2013quantum}%
  \BibitemOpen
  \bibfield  {author} {\bibinfo {author} {\bibfnamefont {R.}~\bibnamefont
  {Willett}},\ }\href@noop {} {\bibfield  {journal} {\bibinfo  {journal}
  {Reports on Progress in Physics}\ }\textbf {\bibinfo {volume} {76}},\
  \bibinfo {pages} {076501} (\bibinfo {year} {2013})}\BibitemShut {NoStop}%
\bibitem [{\citenamefont {Morf}(1998)}]{morf1998transition}%
  \BibitemOpen
  \bibfield  {author} {\bibinfo {author} {\bibfnamefont {R.~H.}\ \bibnamefont
  {Morf}},\ }\href@noop {} {\bibfield  {journal} {\bibinfo  {journal} {Physical
  review letters}\ }\textbf {\bibinfo {volume} {80}},\ \bibinfo {pages} {1505}
  (\bibinfo {year} {1998})}\BibitemShut {NoStop}%
\bibitem [{\citenamefont {Lu}\ \emph {et~al.}(2010)\citenamefont {Lu},
  \citenamefont {Sarma},\ and\ \citenamefont {Park}}]{lu2010superconducting}%
  \BibitemOpen
  \bibfield  {author} {\bibinfo {author} {\bibfnamefont {H.}~\bibnamefont
  {Lu}}, \bibinfo {author} {\bibfnamefont {S.~D.}\ \bibnamefont {Sarma}}, \
  and\ \bibinfo {author} {\bibfnamefont {K.}~\bibnamefont {Park}},\ }\href@noop
  {} {\bibfield  {journal} {\bibinfo  {journal} {Physical Review B}\ }\textbf
  {\bibinfo {volume} {82}},\ \bibinfo {pages} {201303} (\bibinfo {year}
  {2010})}\BibitemShut {NoStop}%
\bibitem [{\citenamefont {Storni}\ \emph {et~al.}(2010)\citenamefont {Storni},
  \citenamefont {Morf},\ and\ \citenamefont {Sarma}}]{storni2010fractional}%
  \BibitemOpen
  \bibfield  {author} {\bibinfo {author} {\bibfnamefont {M.}~\bibnamefont
  {Storni}}, \bibinfo {author} {\bibfnamefont {R.}~\bibnamefont {Morf}}, \ and\
  \bibinfo {author} {\bibfnamefont {S.~D.}\ \bibnamefont {Sarma}},\ }\href@noop
  {} {\bibfield  {journal} {\bibinfo  {journal} {Physical review letters}\
  }\textbf {\bibinfo {volume} {104}},\ \bibinfo {pages} {076803} (\bibinfo
  {year} {2010})}\BibitemShut {NoStop}%
\bibitem [{\citenamefont {Pakrouski}\ \emph {et~al.}(2015)\citenamefont
  {Pakrouski}, \citenamefont {Peterson}, \citenamefont {Jolicoeur},
  \citenamefont {Scarola}, \citenamefont {Nayak},\ and\ \citenamefont
  {Troyer}}]{pakrouski2015phase}%
  \BibitemOpen
  \bibfield  {author} {\bibinfo {author} {\bibfnamefont {K.}~\bibnamefont
  {Pakrouski}}, \bibinfo {author} {\bibfnamefont {M.~R.}\ \bibnamefont
  {Peterson}}, \bibinfo {author} {\bibfnamefont {T.}~\bibnamefont {Jolicoeur}},
  \bibinfo {author} {\bibfnamefont {V.~W.}\ \bibnamefont {Scarola}}, \bibinfo
  {author} {\bibfnamefont {C.}~\bibnamefont {Nayak}}, \ and\ \bibinfo {author}
  {\bibfnamefont {M.}~\bibnamefont {Troyer}},\ }\href@noop {} {\bibfield
  {journal} {\bibinfo  {journal} {Physical Review X}\ }\textbf {\bibinfo
  {volume} {5}},\ \bibinfo {pages} {021004} (\bibinfo {year}
  {2015})}\BibitemShut {NoStop}%
\bibitem [{\citenamefont {Tylan-Tyler}\ and\ \citenamefont
  {Lyanda-Geller}(2015)}]{tylan2015phase}%
  \BibitemOpen
  \bibfield  {author} {\bibinfo {author} {\bibfnamefont {A.}~\bibnamefont
  {Tylan-Tyler}}\ and\ \bibinfo {author} {\bibfnamefont {Y.}~\bibnamefont
  {Lyanda-Geller}},\ }\href@noop {} {\bibfield  {journal} {\bibinfo  {journal}
  {Physical Review B}\ }\textbf {\bibinfo {volume} {91}},\ \bibinfo {pages}
  {205404} (\bibinfo {year} {2015})}\BibitemShut {NoStop}%
\bibitem [{\citenamefont {Banerjee}\ \emph {et~al.}(2017)\citenamefont
  {Banerjee}, \citenamefont {Heiblum}, \citenamefont {Rosenblatt},
  \citenamefont {Oreg}, \citenamefont {Feldman}, \citenamefont {Stern},\ and\
  \citenamefont {Umansky}}]{banerjee2017observed}%
  \BibitemOpen
  \bibfield  {author} {\bibinfo {author} {\bibfnamefont {M.}~\bibnamefont
  {Banerjee}}, \bibinfo {author} {\bibfnamefont {M.}~\bibnamefont {Heiblum}},
  \bibinfo {author} {\bibfnamefont {A.}~\bibnamefont {Rosenblatt}}, \bibinfo
  {author} {\bibfnamefont {Y.}~\bibnamefont {Oreg}}, \bibinfo {author}
  {\bibfnamefont {D.~E.}\ \bibnamefont {Feldman}}, \bibinfo {author}
  {\bibfnamefont {A.}~\bibnamefont {Stern}}, \ and\ \bibinfo {author}
  {\bibfnamefont {V.}~\bibnamefont {Umansky}},\ }\href@noop {} {\bibfield
  {journal} {\bibinfo  {journal} {Nature}\ }\textbf {\bibinfo {volume} {545}},\
  \bibinfo {pages} {75} (\bibinfo {year} {2017})}\BibitemShut {NoStop}%
\bibitem [{\citenamefont {Banerjee}\ \emph {et~al.}(2018)\citenamefont
  {Banerjee}, \citenamefont {Heiblum}, \citenamefont {Umansky}, \citenamefont
  {Feldman}, \citenamefont {Oreg},\ and\ \citenamefont
  {Stern}}]{banerjee2018observation}%
  \BibitemOpen
  \bibfield  {author} {\bibinfo {author} {\bibfnamefont {M.}~\bibnamefont
  {Banerjee}}, \bibinfo {author} {\bibfnamefont {M.}~\bibnamefont {Heiblum}},
  \bibinfo {author} {\bibfnamefont {V.}~\bibnamefont {Umansky}}, \bibinfo
  {author} {\bibfnamefont {D.~E.}\ \bibnamefont {Feldman}}, \bibinfo {author}
  {\bibfnamefont {Y.}~\bibnamefont {Oreg}}, \ and\ \bibinfo {author}
  {\bibfnamefont {A.}~\bibnamefont {Stern}},\ }\href@noop {} {\bibfield
  {journal} {\bibinfo  {journal} {Nature}\ }\textbf {\bibinfo {volume} {559}},\
  \bibinfo {pages} {205} (\bibinfo {year} {2018})}\BibitemShut {NoStop}%
\bibitem [{\citenamefont {Ma}\ \emph {et~al.}(2022)\citenamefont {Ma},
  \citenamefont {Peterson}, \citenamefont {Scarola},\ and\ \citenamefont
  {Yang}}]{ma2022fractional}%
  \BibitemOpen
  \bibfield  {author} {\bibinfo {author} {\bibfnamefont {K.~K.}\ \bibnamefont
  {Ma}}, \bibinfo {author} {\bibfnamefont {M.~R.}\ \bibnamefont {Peterson}},
  \bibinfo {author} {\bibfnamefont {V.}~\bibnamefont {Scarola}}, \ and\
  \bibinfo {author} {\bibfnamefont {K.}~\bibnamefont {Yang}},\ }\href@noop {}
  {\bibfield  {journal} {\bibinfo  {journal} {arXiv preprint arXiv:2208.07908}\
  } (\bibinfo {year} {2022})}\BibitemShut {NoStop}%
\bibitem [{\citenamefont {Francesco}\ \emph {et~al.}(2012)\citenamefont
  {Francesco}, \citenamefont {Mathieu},\ and\ \citenamefont
  {S{\'e}n{\'e}chal}}]{francesco2012conformal}%
  \BibitemOpen
  \bibfield  {author} {\bibinfo {author} {\bibfnamefont {P.}~\bibnamefont
  {Francesco}}, \bibinfo {author} {\bibfnamefont {P.}~\bibnamefont {Mathieu}},
  \ and\ \bibinfo {author} {\bibfnamefont {D.}~\bibnamefont
  {S{\'e}n{\'e}chal}},\ }\href@noop {} {\emph {\bibinfo {title} {Conformal
  field theory}}}\ (\bibinfo  {publisher} {Springer Science \& Business
  Media},\ \bibinfo {year} {2012})\BibitemShut {NoStop}%
\bibitem [{\citenamefont {Cano}\ \emph {et~al.}(2014)\citenamefont {Cano},
  \citenamefont {Cheng}, \citenamefont {Mulligan}, \citenamefont {Nayak},
  \citenamefont {Plamadeala},\ and\ \citenamefont {Yard}}]{cano2014bulk}%
  \BibitemOpen
  \bibfield  {author} {\bibinfo {author} {\bibfnamefont {J.}~\bibnamefont
  {Cano}}, \bibinfo {author} {\bibfnamefont {M.}~\bibnamefont {Cheng}},
  \bibinfo {author} {\bibfnamefont {M.}~\bibnamefont {Mulligan}}, \bibinfo
  {author} {\bibfnamefont {C.}~\bibnamefont {Nayak}}, \bibinfo {author}
  {\bibfnamefont {E.}~\bibnamefont {Plamadeala}}, \ and\ \bibinfo {author}
  {\bibfnamefont {J.}~\bibnamefont {Yard}},\ }\href@noop {} {\bibfield
  {journal} {\bibinfo  {journal} {Physical Review B}\ }\textbf {\bibinfo
  {volume} {89}},\ \bibinfo {pages} {115116} (\bibinfo {year}
  {2014})}\BibitemShut {NoStop}%
\bibitem [{\citenamefont {Maslov}\ and\ \citenamefont
  {Stone}(1995)}]{maslov1995landauer}%
  \BibitemOpen
  \bibfield  {author} {\bibinfo {author} {\bibfnamefont {D.~L.}\ \bibnamefont
  {Maslov}}\ and\ \bibinfo {author} {\bibfnamefont {M.}~\bibnamefont {Stone}},\
  }\href@noop {} {\bibfield  {journal} {\bibinfo  {journal} {Physical Review
  B}\ }\textbf {\bibinfo {volume} {52}},\ \bibinfo {pages} {R5539} (\bibinfo
  {year} {1995})}\BibitemShut {NoStop}%
\bibitem [{\citenamefont {Moore}\ and\ \citenamefont
  {Read}(1991)}]{moore1991nonabelions}%
  \BibitemOpen
  \bibfield  {author} {\bibinfo {author} {\bibfnamefont {G.}~\bibnamefont
  {Moore}}\ and\ \bibinfo {author} {\bibfnamefont {N.}~\bibnamefont {Read}},\
  }\href@noop {} {\bibfield  {journal} {\bibinfo  {journal} {Nuclear Physics
  B}\ }\textbf {\bibinfo {volume} {360}},\ \bibinfo {pages} {362} (\bibinfo
  {year} {1991})}\BibitemShut {NoStop}%
\bibitem [{\citenamefont {Son}(2015)}]{son2015composite}%
  \BibitemOpen
  \bibfield  {author} {\bibinfo {author} {\bibfnamefont {D.~T.}\ \bibnamefont
  {Son}},\ }\href@noop {} {\bibfield  {journal} {\bibinfo  {journal} {Physical
  Review X}\ }\textbf {\bibinfo {volume} {5}},\ \bibinfo {pages} {031027}
  (\bibinfo {year} {2015})}\BibitemShut {NoStop}%
\bibitem [{\citenamefont {Jain}(1989)}]{jain1989incompressible}%
  \BibitemOpen
  \bibfield  {author} {\bibinfo {author} {\bibfnamefont {J.~K.}\ \bibnamefont
  {Jain}},\ }\href@noop {} {\bibfield  {journal} {\bibinfo  {journal} {Physical
  Review B}\ }\textbf {\bibinfo {volume} {40}},\ \bibinfo {pages} {8079}
  (\bibinfo {year} {1989})}\BibitemShut {NoStop}%
\bibitem [{\citenamefont {Levin}\ \emph {et~al.}(2007)\citenamefont {Levin},
  \citenamefont {Halperin},\ and\ \citenamefont {Rosenow}}]{levin2007particle}%
  \BibitemOpen
  \bibfield  {author} {\bibinfo {author} {\bibfnamefont {M.}~\bibnamefont
  {Levin}}, \bibinfo {author} {\bibfnamefont {B.~I.}\ \bibnamefont {Halperin}},
  \ and\ \bibinfo {author} {\bibfnamefont {B.}~\bibnamefont {Rosenow}},\
  }\href@noop {} {\bibfield  {journal} {\bibinfo  {journal} {Physical review
  letters}\ }\textbf {\bibinfo {volume} {99}},\ \bibinfo {pages} {236806}
  (\bibinfo {year} {2007})}\BibitemShut {NoStop}%
\bibitem [{\citenamefont {Lee}\ \emph {et~al.}(2007)\citenamefont {Lee},
  \citenamefont {Ryu}, \citenamefont {Nayak},\ and\ \citenamefont
  {Fisher}}]{lee2007particle}%
  \BibitemOpen
  \bibfield  {author} {\bibinfo {author} {\bibfnamefont {S.-S.}\ \bibnamefont
  {Lee}}, \bibinfo {author} {\bibfnamefont {S.}~\bibnamefont {Ryu}}, \bibinfo
  {author} {\bibfnamefont {C.}~\bibnamefont {Nayak}}, \ and\ \bibinfo {author}
  {\bibfnamefont {M.~P.}\ \bibnamefont {Fisher}},\ }\href@noop {} {\bibfield
  {journal} {\bibinfo  {journal} {Physical review letters}\ }\textbf {\bibinfo
  {volume} {99}},\ \bibinfo {pages} {236807} (\bibinfo {year}
  {2007})}\BibitemShut {NoStop}%
\bibitem [{\citenamefont {Wen}(2004)}]{wen2004quantum}%
  \BibitemOpen
  \bibfield  {author} {\bibinfo {author} {\bibfnamefont {X.-G.}\ \bibnamefont
  {Wen}},\ }\href@noop {} {\emph {\bibinfo {title} {Quantum field theory of
  many-body systems: from the origin of sound to an origin of light and
  electrons}}}\ (\bibinfo  {publisher} {OUP Oxford},\ \bibinfo {year}
  {2004})\BibitemShut {NoStop}%
\bibitem [{Note1()}]{Note1}%
  \BibitemOpen
  \bibinfo {note} {Here we are leaving the Klein factors needed for the fermion
  anti-commutation relationships implicit}\BibitemShut {NoStop}%
\bibitem [{\citenamefont {Milovanovi{\'c}}\ and\ \citenamefont
  {Read}(1996)}]{milovanovic1996edge}%
  \BibitemOpen
  \bibfield  {author} {\bibinfo {author} {\bibfnamefont {M.}~\bibnamefont
  {Milovanovi{\'c}}}\ and\ \bibinfo {author} {\bibfnamefont {N.}~\bibnamefont
  {Read}},\ }\href@noop {} {\bibfield  {journal} {\bibinfo  {journal} {Physical
  Review B}\ }\textbf {\bibinfo {volume} {53}},\ \bibinfo {pages} {13559}
  (\bibinfo {year} {1996})}\BibitemShut {NoStop}%
\bibitem [{\citenamefont {Sohal}\ \emph {et~al.}(2020)\citenamefont {Sohal},
  \citenamefont {Han}, \citenamefont {Santos},\ and\ \citenamefont
  {Teo}}]{sohal2020entanglement}%
  \BibitemOpen
  \bibfield  {author} {\bibinfo {author} {\bibfnamefont {R.}~\bibnamefont
  {Sohal}}, \bibinfo {author} {\bibfnamefont {B.}~\bibnamefont {Han}}, \bibinfo
  {author} {\bibfnamefont {L.~H.}\ \bibnamefont {Santos}}, \ and\ \bibinfo
  {author} {\bibfnamefont {J.~C.}\ \bibnamefont {Teo}},\ }\href@noop {}
  {\bibfield  {journal} {\bibinfo  {journal} {Physical Review B}\ }\textbf
  {\bibinfo {volume} {102}},\ \bibinfo {pages} {045102} (\bibinfo {year}
  {2020})}\BibitemShut {NoStop}%
\bibitem [{\citenamefont {Lim}\ \emph {et~al.}(2021)\citenamefont {Lim},
  \citenamefont {Asasi}, \citenamefont {Teo},\ and\ \citenamefont
  {Mulligan}}]{lim2021disentangling}%
  \BibitemOpen
  \bibfield  {author} {\bibinfo {author} {\bibfnamefont {P.~K.}\ \bibnamefont
  {Lim}}, \bibinfo {author} {\bibfnamefont {H.}~\bibnamefont {Asasi}}, \bibinfo
  {author} {\bibfnamefont {J.~C.}\ \bibnamefont {Teo}}, \ and\ \bibinfo
  {author} {\bibfnamefont {M.}~\bibnamefont {Mulligan}},\ }\href@noop {}
  {\bibfield  {journal} {\bibinfo  {journal} {Physical Review B}\ }\textbf
  {\bibinfo {volume} {104}},\ \bibinfo {pages} {115155} (\bibinfo {year}
  {2021})}\BibitemShut {NoStop}%
\bibitem [{Note2()}]{Note2}%
  \BibitemOpen
  \bibinfo {note} {We will only consider systems where the bulk topological
  order is defined on a manifold with a single edge, (e.g., discs). If the
  manifold has multiple decoupled edges, (e.g., a corbino disk or a cylinder)
  only the net parity of all edges is conserved}\BibitemShut {NoStop}%
\bibitem [{\citenamefont {Gross}\ and\ \citenamefont
  {Neveu}(1974)}]{gross1974dynamical}%
  \BibitemOpen
  \bibfield  {author} {\bibinfo {author} {\bibfnamefont {D.~J.}\ \bibnamefont
  {Gross}}\ and\ \bibinfo {author} {\bibfnamefont {A.}~\bibnamefont {Neveu}},\
  }\href@noop {} {\bibfield  {journal} {\bibinfo  {journal} {Physical Review
  D}\ }\textbf {\bibinfo {volume} {10}},\ \bibinfo {pages} {3235} (\bibinfo
  {year} {1974})}\BibitemShut {NoStop}%
\bibitem [{\citenamefont {Witten}(1978)}]{witten1978some}%
  \BibitemOpen
  \bibfield  {author} {\bibinfo {author} {\bibfnamefont {E.}~\bibnamefont
  {Witten}},\ }\href@noop {} {\bibfield  {journal} {\bibinfo  {journal}
  {Nuclear Physics B}\ }\textbf {\bibinfo {volume} {142}},\ \bibinfo {pages}
  {285} (\bibinfo {year} {1978})}\BibitemShut {NoStop}%
\bibitem [{\citenamefont {Zamolodchikov}\ and\ \citenamefont
  {Zamolodchikov}(1978)}]{zamolodchikov1978exact}%
  \BibitemOpen
  \bibfield  {author} {\bibinfo {author} {\bibfnamefont {A.~B.}\ \bibnamefont
  {Zamolodchikov}}\ and\ \bibinfo {author} {\bibfnamefont {A.~B.}\ \bibnamefont
  {Zamolodchikov}},\ }\href@noop {} {\bibfield  {journal} {\bibinfo  {journal}
  {Physics Letters B}\ }\textbf {\bibinfo {volume} {72}},\ \bibinfo {pages}
  {481} (\bibinfo {year} {1978})}\BibitemShut {NoStop}%
\bibitem [{\citenamefont {Fendley}\ and\ \citenamefont
  {Saleur}(2001)}]{fendley2001bps}%
  \BibitemOpen
  \bibfield  {author} {\bibinfo {author} {\bibfnamefont {P.}~\bibnamefont
  {Fendley}}\ and\ \bibinfo {author} {\bibfnamefont {H.}~\bibnamefont
  {Saleur}},\ }\href@noop {} {\bibfield  {journal} {\bibinfo  {journal}
  {Physical Review D}\ }\textbf {\bibinfo {volume} {65}},\ \bibinfo {pages}
  {025001} (\bibinfo {year} {2001})}\BibitemShut {NoStop}%
\bibitem [{\citenamefont {Clarke}\ \emph {et~al.}(2013)\citenamefont {Clarke},
  \citenamefont {Alicea},\ and\ \citenamefont {Shtengel}}]{clarke2013exotic}%
  \BibitemOpen
  \bibfield  {author} {\bibinfo {author} {\bibfnamefont {D.~J.}\ \bibnamefont
  {Clarke}}, \bibinfo {author} {\bibfnamefont {J.}~\bibnamefont {Alicea}}, \
  and\ \bibinfo {author} {\bibfnamefont {K.}~\bibnamefont {Shtengel}},\
  }\href@noop {} {\bibfield  {journal} {\bibinfo  {journal} {Nature
  communications}\ }\textbf {\bibinfo {volume} {4}},\ \bibinfo {pages} {1348}
  (\bibinfo {year} {2013})}\BibitemShut {NoStop}%
\bibitem [{\citenamefont {Alicea}\ and\ \citenamefont
  {Fendley}(2016)}]{alicea2016topological}%
  \BibitemOpen
  \bibfield  {author} {\bibinfo {author} {\bibfnamefont {J.}~\bibnamefont
  {Alicea}}\ and\ \bibinfo {author} {\bibfnamefont {P.}~\bibnamefont
  {Fendley}},\ }\href@noop {} {\bibfield  {journal} {\bibinfo  {journal}
  {Annual Review of Condensed Matter Physics}\ }\textbf {\bibinfo {volume}
  {7}},\ \bibinfo {pages} {119} (\bibinfo {year} {2016})}\BibitemShut {NoStop}%
\bibitem [{\citenamefont {Jackiw}\ and\ \citenamefont
  {Rebbi}(1976)}]{jackiw1976solitons}%
  \BibitemOpen
  \bibfield  {author} {\bibinfo {author} {\bibfnamefont {R.}~\bibnamefont
  {Jackiw}}\ and\ \bibinfo {author} {\bibfnamefont {C.}~\bibnamefont {Rebbi}},\
  }\href@noop {} {\bibfield  {journal} {\bibinfo  {journal} {Physical Review
  D}\ }\textbf {\bibinfo {volume} {13}},\ \bibinfo {pages} {3398} (\bibinfo
  {year} {1976})}\BibitemShut {NoStop}%
\bibitem [{\citenamefont {Fu}\ and\ \citenamefont
  {Kane}(2008)}]{fu2008superconducting}%
  \BibitemOpen
  \bibfield  {author} {\bibinfo {author} {\bibfnamefont {L.}~\bibnamefont
  {Fu}}\ and\ \bibinfo {author} {\bibfnamefont {C.~L.}\ \bibnamefont {Kane}},\
  }\href@noop {} {\bibfield  {journal} {\bibinfo  {journal} {Physical review
  letters}\ }\textbf {\bibinfo {volume} {100}},\ \bibinfo {pages} {096407}
  (\bibinfo {year} {2008})}\BibitemShut {NoStop}%
\bibitem [{\citenamefont {Zhang}(2024)}]{zhang2024vortex}%
  \BibitemOpen
  \bibfield  {author} {\bibinfo {author} {\bibfnamefont {Y.-H.}\ \bibnamefont
  {Zhang}},\ }\href@noop {} {\bibfield  {journal} {\bibinfo  {journal} {arXiv
  preprint arXiv:2402.05112}\ } (\bibinfo {year} {2024})}\BibitemShut {NoStop}%
\bibitem [{\citenamefont {Shi}\ \emph {et~al.}(2024)\citenamefont {Shi},
  \citenamefont {Goldman}, \citenamefont {Dong},\ and\ \citenamefont
  {Senthil}}]{shi2024excitonic}%
  \BibitemOpen
  \bibfield  {author} {\bibinfo {author} {\bibfnamefont {Z.~D.}\ \bibnamefont
  {Shi}}, \bibinfo {author} {\bibfnamefont {H.}~\bibnamefont {Goldman}},
  \bibinfo {author} {\bibfnamefont {Z.}~\bibnamefont {Dong}}, \ and\ \bibinfo
  {author} {\bibfnamefont {T.}~\bibnamefont {Senthil}},\ }\href@noop {}
  {\bibfield  {journal} {\bibinfo  {journal} {arXiv preprint arXiv:2402.12436}\
  } (\bibinfo {year} {2024})}\BibitemShut {NoStop}%
\bibitem [{\citenamefont {Shytov}\ \emph {et~al.}(1998)\citenamefont {Shytov},
  \citenamefont {Levitov},\ and\ \citenamefont
  {Halperin}}]{shytov1998tunneling}%
  \BibitemOpen
  \bibfield  {author} {\bibinfo {author} {\bibfnamefont {A.}~\bibnamefont
  {Shytov}}, \bibinfo {author} {\bibfnamefont {L.}~\bibnamefont {Levitov}}, \
  and\ \bibinfo {author} {\bibfnamefont {B.}~\bibnamefont {Halperin}},\
  }\href@noop {} {\bibfield  {journal} {\bibinfo  {journal} {Physical review
  letters}\ }\textbf {\bibinfo {volume} {80}},\ \bibinfo {pages} {141}
  (\bibinfo {year} {1998})}\BibitemShut {NoStop}%
\bibitem [{\citenamefont {Levitov}\ \emph {et~al.}(2001)\citenamefont
  {Levitov}, \citenamefont {Shytov},\ and\ \citenamefont
  {Halperin}}]{levitov2001effective}%
  \BibitemOpen
  \bibfield  {author} {\bibinfo {author} {\bibfnamefont {L.}~\bibnamefont
  {Levitov}}, \bibinfo {author} {\bibfnamefont {A.}~\bibnamefont {Shytov}}, \
  and\ \bibinfo {author} {\bibfnamefont {B.}~\bibnamefont {Halperin}},\
  }\href@noop {} {\bibfield  {journal} {\bibinfo  {journal} {Physical Review
  B}\ }\textbf {\bibinfo {volume} {64}},\ \bibinfo {pages} {075322} (\bibinfo
  {year} {2001})}\BibitemShut {NoStop}%
\bibitem [{\citenamefont {Barkeshli}\ \emph {et~al.}(2015)\citenamefont
  {Barkeshli}, \citenamefont {Mulligan},\ and\ \citenamefont
  {Fisher}}]{barkeshli2015particle}%
  \BibitemOpen
  \bibfield  {author} {\bibinfo {author} {\bibfnamefont {M.}~\bibnamefont
  {Barkeshli}}, \bibinfo {author} {\bibfnamefont {M.}~\bibnamefont {Mulligan}},
  \ and\ \bibinfo {author} {\bibfnamefont {M.~P.}\ \bibnamefont {Fisher}},\
  }\href@noop {} {\bibfield  {journal} {\bibinfo  {journal} {Physical Review
  B}\ }\textbf {\bibinfo {volume} {92}},\ \bibinfo {pages} {165125} (\bibinfo
  {year} {2015})}\BibitemShut {NoStop}%
\end{thebibliography}%
\bibliographystyle{apsrev4-1}

\appendix 
\section{Interacting Edges of Abelian FQSHs}\label{app:AbelainIntEdge}
Here we consider the edge theory of a generic Abelian FQSH with U$(1)$ charge conservation, $s^z$-spin conservation and time-reversal symmetry. We are mainly interested in systems where the spin-Hall conductance is a half-integer in this appendix, but the results here apply to arbitrary Abelian FQSHs. The Abelian FQSH edge theory can be written in terms of $2N$ bosonic degrees of freedom, $N$ per spin. The bosonic edge Lagrangian in the K-matrix form is
\begin{equation}\begin{split}
\mathcal{L} = &\tfrac{1}{4\pi} \partial_t \bm{\Phi}^T \bm{K}  \partial_x  \bm{\Phi} - \tfrac{1}{4\pi} \partial_x  \bm{\Phi}^T \bm{V} \partial_x \bm{\Phi} \\[1pt]
&+ \tfrac{1}{2\pi} \epsilon^{\mu\nu} \partial_\mu (\bm{t}\cdot \bm{\Phi}) A_\nu + \tfrac{1}{2\pi} \epsilon^{\mu\nu} \partial_\mu (\bm{s}\cdot \bm{\Phi})  A^{\text{s}}_\nu.
\label{eq:AppBosonLagrangian}\end{split}\end{equation}    
where $\bm{K}$ is a $2N \times 2N$ integer valued symmetric matrix, $\bm{V}$ is the velocity matrix, and $\bm{\Phi}$, $\bm{t}$, and $\bm{s}$ are the bosonic, charge, and spin vectors respectively. In the last line of Eq.~\ref{eq:AppBosonLagrangian}, we have included couplings to probe  electromagnetic U$(1)$ gauge field $A$, and a probe spin U$(1)_\text{s}$ gauge probe $A^{\text{s}}$. The anyons of the edge correspond to vertex operators, $e^{i \bm{l}\cdot \bm{\Phi}}$ where $\bm{l}$ is a $2N$-component integer vector. The commutation relationship for bosons $\bm{l}\cdot \bm{\Phi}$ and $\bm{l}' \cdot \bm{\Phi}$ is $[\bm{l}\cdot \bm{\Phi}(x), \bm{l}' \cdot \bm{\Phi}(x')] = \pi \bm{l}^T \bm{K}^{-1} \bm{l}' \text{sgn}(x-x')$. 
The quantum Hall (QH), and quantum-spin Hall (QSH) conductances are given by $\sigma_H = \bm{t}^T \bm{K}^{-1} \bm{t}$, and $\sigma_{sH} = \bm{s}^T \bm{K}^{-1} \bm{t}$  respectively. For a time-reversal symmetric system $\sigma_H = 0$. Without loss of generality, we take $\sigma_{sH} > 0$ such that the spin-up modes carry charge upstream (i.e., the $+x$-direction).

For a fermionic system with charge conservation, $s^z$-conservation, and TRS, the matrix $\bm{K}$ can be written as
\begin{equation}\begin{split}
&\phantom{====}\bm{K} = \begin{bmatrix} \bm{K}_\uparrow  & \bm{0}_{N\times N}\\
 \bm{0}_{N\times N} & \bm{K}_\downarrow  \end{bmatrix}
 \end{split}\end{equation}
 with $\bm{t} = \bm{1}_{2N}$, $\bm{s}  = \frac{1}{2}( \bm{1}_{N}, -\bm{1}_{N})$ where $\bm{1}_{N}$ is the $N$ component vector of all ones. We label the bosonic fields as $\bm{\Phi}  = (\phi_{\uparrow 1},..., \phi_{\uparrow N}, \phi_{\downarrow 1},..., \phi_{\downarrow N})$. The $N\times N$ matrix blocks satisfy $\bm{K}_\uparrow = -\bm{K}_\downarrow$. TRS acts on the bosons as
$\phi_{\uparrow n} \rightarrow  \phi_{\downarrow n}$, $\phi_{\uparrow n} \rightarrow  \phi_{\uparrow n} + \pi [\bm{e}^T_{n} \bm{K}^{-1} \bm{t} ] $
where $1\leq n \leq N$, and $\bm{e}^T_{m}$ is the $2N$-component vector where the $m^\text{th}$ element is $1$ and all other elements are $0$. 

The local electrons are given by vertex operators 
\begin{equation}\begin{split}
\Psi^\dagger_{\uparrow n} &= e^{i \bm{e}^T_{n} \bm{K} \bm{\Phi}},\phantom{=}
\Psi^\dagger_{\downarrow n} = e^{i \bm{e}^T_{N+n} \bm{K} \bm{\Phi}}.
\label{eq:AbelainElectronsDef}\end{split}\end{equation}
These operators carry charge $1$ and spin $\pm \frac{1}{2}$. Furthermore, the electrons form Kramer's pairs under TRS, $\Psi^\dagger_{\uparrow n} \rightarrow \Psi^\dagger_{\downarrow n}$ $\Psi^\dagger_{\downarrow n} \rightarrow -\Psi^\dagger_{\uparrow n}$. In principle, it is also necessary to include Klein factors in this definition, such that the different fermions anti-commute with each other. However, the inclusion of these terms does not change our results, and so we will omit them for brevity.

The quadratic bosonic Lagrangian in Eq~\ref{eq:AppBosonLagrangian} can be supplemented with additional interactions. In a local theory, any interactions must be written in terms of the local electron operators in Eq.~\ref{eq:AbelainElectronsDef}. Here, we consider the following double-spin flip operations,
\begin{equation}\begin{split}
\Psi^\dagger_{\uparrow 1} \Psi_{\uparrow n} \Psi^\dagger_{\downarrow n} \Psi_{\downarrow 1} + h.c., = 2\cos( \bm{\Lambda}^T_{n} \bm{K} \bm{\Phi} ),
\label{eq:AbelianInteractions}\end{split}\end{equation}
for $2 \leq n \leq N$. Here $\bm{\Lambda}_{n} = \bm{e}_{1} - \bm{e}_{n} - \bm{e}_{N+1} + \bm{e}_{N+n}$. These interactions conserve charge and $s^z$, and are time-reversal symmetric. As is usual in K-matrix theories, these interactions are relevant for an appropriate choice of the velocity matrix $\bm{V}$. 

At strong coupling, these interactions will gap out all the bosons, except for two. This can be directly confirmed by noting that the $N-1$ linearly independent vectors $\bm{\Lambda}_{n}$ satisfy the null vector criteria $\bm{\Lambda}_{n} \bm{K} \bm{\Lambda}_{n'} = 0$. Each null vector will remove two bosonic fields from the spectrum, the field in the cosine terms, and its dual field. The reamining two gapless bosons are 
\begin{equation}\begin{split}
\phi_{R} = [\tfrac{1}{2}\bm{t} + \bm{s}]\cdot \bm{\Phi}, \phantom{=} \phi_{L} = [\tfrac{1}{2}\bm{t} - \bm{s}]\cdot \bm{\Phi}.
\label{eq:gaplessBosons}\end{split}\end{equation}
These bosons commute will all the interactions in Eq.~\ref{eq:AbelianInteractions}. They are also not condensed by the cosine terms, since $\phi_R$ and $\phi_L$ carry spin and charge, while all the condensed bosons are charge and spin neutral. $\phi_R$ and $\phi_L$ are therefore non-trivial gapless degrees of freedom at the edge. The Lagrangian for these bosons is Eq.~\ref{eq:effectiveLag} with $\sigma_{sH} = \bm{t}^T \bm{K}\bm{s}$. 

Adding additional interactions of the form $\Psi^\dagger_{\uparrow n} \Psi_{\uparrow n'} \Psi^\dagger_{\downarrow n'} \Psi_{\downarrow n}$ to Eq.~\ref{eq:AbelianInteractions} does not change this result. As discussed in the main text, the effective edge theory is stable, and cannot be gapped out without breaking charge or spin conservation. 

Finally,  we would like to determine the local operators in the theory. Specifically, we are looking for the lowest non-zero $p \in \mathbb{Z}$ such that $\exp(i p \phi_{R})$ is a local operator. In the K-matrix formalism, this means that $\exp(i p \phi_{R}) = \exp(i \bm{l}^T\bm{K}\bm{\Phi})$ for some integer valued vector $\bm{l}$. Using that $\phi_{R} = (\frac{1}{2}\bm{t} + \bm{s})\cdot \bm{\Phi}$, we have that $p \bm{K}^{-1}(\frac{1}{2}\bm{t} + \bm{s}) = \bm{l}$, and 
\begin{equation}
    p [\bm{e}^T_{n} \bm{K}^{-1}(\tfrac{1}{2}\bm{t} + \bm{s})] = \bm{e}_{n}\cdot\bm{l} \in \mathbb{Z}.
\end{equation}
Since $\frac{1}{2}\bm{t} + \bm{s} = (\bm{1}_{N}, \bm{0}_{N})$, where $\bm{0}_{N}$ is a $N$-component vector of zeros, $\bm{e}^T_{n} \bm{K}^{-1}(\frac{1}{2}\bm{t} + \bm{s})$ is the electric charge of the vertex operator $\exp(i \phi_{\uparrow n})$ for $1\leq n \leq N$. 

Based on this, the lowest $p$ such that $\exp(i p \phi_{R})$ is local is also the lowest $p$ such that $p$ times any anyon charge is an integer. In a topologically ordered system, every anyon charge is an integer multiple of the lowest non-zero anyon charge $e^*$. The lowest $p$ such that  $\exp(i p \phi_{R})$ is local is therefore $p =  1/e^*$ where $e^*$ is the lowest anyon charge in units of the electron charge $e$. The local operators in the theory are then $\exp(i  \phi_{R}/e^*)$, $\exp(-i  \phi_{L}/e^*)$, and fusions thereof. The main cases of interest in this appendix are the states with half-integer spin Hall conductance. In all these theories, the lowest anyon charge $1/4$\cite{levin2009fractional}. The local operators in such a theory are $\exp(i 4 \phi_{R})$, and $\exp(-i 4 \phi_{L})$. These operators carry charge $4\sigma_{sH}$, indicating that they are combinations of $4\sigma_{sH}$-electron operators. Since $4\sigma_{sH}\in 2 \mathbb{Z}$, these operators are all bosonic. In the simplest case of $\sigma_{sH} = \frac{1}{2}$, both $\exp(i 4 \phi_{R})$, and $\exp(-i 4 \phi_{L})$ are two electron operators.

\section{Scaling dimensions of interactions}\label{app:RGFlows}
In this section, we will discuss the scaling dimensions of the various operators considered in the main text. 

First, we consider the Gross-Neveu interaction Eq.~\ref{eq:GrossNeveu}. This interaction is marginally relevant for $J < 0$ with $\beta$ function, $a \frac{d J}{da } = - \frac{J^2}{4\pi}$, representing a weak coupling instability. Here $J < 0$ corresponds to an anti-ferromagnetic coupling between the up and down spins. It is also possible to make this term relevant if we break the $O(M)$ symmetry of the Majorana sector. Let us consider this for $M = 3$ for simplicity, and bosonize the complex fermions, $\psi_{\uparrow 2} + i \psi_{\uparrow 3} = \exp(i \phi'_{\uparrow})$ and $\psi_{\downarrow 2} - i \psi_{\downarrow 3} = \exp(-i \phi'_{\downarrow})$. In terms of the new bosons, the Gross-Neveu interaction is\cite{witten1978some}
\begin{equation}\begin{split}
    \frac{J}{2}&(\sum^3_{m=1} i\psi_{\uparrow m}\psi_{\downarrow m})^2  \\ &= J i\psi_{\uparrow 1}\psi_{\downarrow 1} \cos(\varphi') + \frac{J}
{2}\cos(\varphi')^2,
\end{split}\end{equation}
where we have defined the non-chiral boson $\varphi' = \phi'_{\uparrow} + \phi'_{\downarrow}$ with Luttinger factor $\kappa'$. The two operators have scaling dimension $1 + \kappa'$ and $2\kappa'$ respectively. O$(3)$ symmetry requires that $\kappa' = 1$, making both terms marginal. However, if O$(3)$ symmetry is broken, we can set $\kappa' < 1$ making both terms relevant, independent of the sign of $J$. 

Second, we consider the interactions in Eqs.~\ref{eq:1MgapTerm}, \ref{eq:DoubleSpinFlip}, \ref{eq:TRSSInt1} and \ref{eq:TRSSpinFlip}. For simplicity, we will assume that $\bm{V}$ is chosen such that it only couples fields of the same $\alpha$, with $\alpha=P,1,..$, i.e., its terms are of the form 
$\partial_x \phi_{\uparrow \alpha} \partial_x \phi_{\downarrow \alpha}$, $\partial_x \phi_{\uparrow \alpha} \partial_x \phi_{\uparrow \alpha}$ or $\partial_x \phi_{\downarrow \alpha} \partial_x \phi_{\downarrow \alpha}$ for $\alpha = P, 1, 2,...$. In this case the velocity matrix $\bm{V}$ is block diagonal, and we can define the Luttinger parameters $\kappa_{\alpha}$ of the boson $\varphi_{\alpha} = \phi_{\uparrow \alpha} + \phi_{\downarrow \alpha}$. The terms we have neglected here lead to more complicated scaling relationships for the bosonic vertex operators. It is straightforward (albeit tedious) to include these terms. 

The scaling dimension of the two terms in Eq.~\ref{eq:1MgapTerm} are $1 + 4 \kappa_{P} + \kappa_1$ and $1 + 4 \kappa_{P} + \kappa_2$. If $4 \kappa_{P} + \kappa_1 < 1$, and $4 \kappa_{P} + \kappa_2 < 1$ then the operators are relevant. This condition can be satisfied by adding repulsive density-density interactions of the form $\chi^\dagger_{\uparrow 1}\chi_{\uparrow 1} \chi^\dagger_{\downarrow 1}\chi_{\downarrow 1}$, and $\Psi^\dagger_{\uparrow n}\Psi_{\uparrow n} \Psi^\dagger_{\downarrow n}\Psi_{\downarrow n}$ for $n = 1 ,2$. Notably, repulsive interactions are required for the 1DQW electrons, as these are not present, $\kappa_{1} = \kappa_{2} = 1$, and the interaction is irrelevant for any value of $\kappa_{P}$. The remaining edge interactions we consider in the main text have similar forms and analysis to those listed above. The interactions in Eq.~\ref{eq:DoubleSpinFlip} have scaling dimension $1 + 4 \kappa_{P} + \kappa_n$, and are relevant when  $4 \kappa_{P} + \kappa_n < 1$. The interaction in Eq.~\ref{eq:TRSSInt1} is purely bosonic  and has scaling dimension $8 \kappa_{P}$. It is relevant when $\kappa_P < \frac{1}{4}$. The interactions in Eq.~\ref{eq:TRSSpinFlip} has scaling dimension $1 + 4 \kappa_{P}$, and is also relevant when $\kappa_P < \frac{1}{4}$.

\section{Mass generation from Majorana bilinears and bosonic vertex operators}\label{app:MajAndVertex}
In this section we will discuss mass generation from terms of the form
\begin{equation}
    H_{int} = g i \psi_{R} \psi_{L}\cos(\varphi)
\label{eq:AppMajVertexInt}\end{equation}
where $H_{int}$ is an Hamiltonian density, $\psi_{R}(x)$ and $\psi_{L}(x)$ are left and right moving Majorana fermions, and $\varphi(x)$ is some non-chiral boson that commutes with itself at different positions. If the Luttinger parameter of $\varphi$ is $\kappa$, then this term has scaling dimension $1 + \kappa$, and is therefore relevant for $\kappa < 1$. When this term is relevant, and flows to strong coupling, we assume that all bosonic operators take on classical expectation values. This motivates the following mean field decomposition\cite{sohal2020entanglement, lim2021disentangling},
\begin{equation}
    H_{int} = g i \psi_{R} \psi_{L}\cos(\varphi) \rightarrow  m i \psi_{\uparrow} \psi_{\downarrow} + h \cos(\varphi)
\end{equation}
where $m = g \langle \cos(\varphi) \rangle $ and $h = g \langle i \psi_{\uparrow} \psi_{\downarrow}  \rangle $. In the strong coupling limit, $g\rightarrow \pm \infty$, and $\kappa \rightarrow 0$ such that $g$ has scaling dimension $1$. In the strong coupling limit, the self-consistent solutions are $\langle \cos(\varphi) \rangle = \pm 1$, $\langle i \psi_{\uparrow} \psi_{\downarrow}  \rangle = \pm \pi \frac{|g|}{v_f} \log( \frac{\Lambda^2}{v_f^2 g^2})$, where $\Lambda$ is the UV cutoff for the Majorana fermions. The expectation value of the Majorana bilinear and the cosine term must have the same sign if $g$ is negative, and opposite signs if $g$ is positive. 

Based on this logic, if we were considering a theory with a Hilbert space that were spanned by the Majorana and bosonic sectors, then this term will have two degenerate ground states. However, for the systems considered in the main text, there is a parity constraint which restricts the Hilbert space. The parity constraint flips the sign of one of the Majoranas and shifts the phase of one of the bosons. For the Majoranas and bosons used in Eq.~\ref{eq:AppMajVertexInt}, the two parity transformation that leaves the physical Hilbert space invariant are $\psi_{R} \rightarrow - \psi_{R}$, $\varphi \rightarrow \varphi + \pi$, and $\psi_{L} \rightarrow - \psi_{L}$, $\varphi \rightarrow \varphi + \pi$ (individually for the left and right Majoranas). 

With this in mind, let us consider the ground state of Eq.~\ref{eq:AppMajVertexInt} in the physical Hilbert space that is invariant under the parity transformation. We begin by assuming $g < 0$ in Eq.~\ref{eq:AppMajVertexInt} (the $g>0$ interaction can be understood analogously). With this interaction, there are two degenerate ground states in the expanded Hilbert space (i.e., the Hilbert space that is spanned by the bosonic and Majorana sectors individually). We label these degenerate states $\ket{+,+}$ and $\ket{-,-}$ where $\langle i \psi_{\uparrow} \psi_{\downarrow}  \rangle>0$, and  $\langle \cos(\varphi) \rangle > 0$ in  $\ket{+,+}$ and $\langle i \psi_{\uparrow} \psi_{\downarrow}  \rangle < 0$, and  $\langle \cos(\varphi) \rangle < 0$ in $\ket{-,-}$. Based on this, the both of the two parity transformation exchange $\ket{+,+}$ and $\ket{-,-}$.
The ground state in the physical Hilbert space corresponds to the superposition of the two degenerate ground states in the expanded Hilbert space $\frac{1}{\sqrt{2}}(\ket{+,+} + \ket{-,-})$. 

In the physical Hilbert space, the Majorana bilinear and cosine terms both do not have expectation values, reflecting the fact that they are not invariant under the parity transformation and are therefore non-local. However, the combination of the terms $i \psi_{R} \psi_{L}\cos(\varphi)$ has a non-vanishing expectation value as it is local and invariant under the parity transformation. Note that although the Majorana bilinear and cosine term no longer have expectation values in the physical Hilbert space, the Majoranas and boson $\varphi$ are still operatots that couple only to gapped excitations. This can be understood by first observing  that the Majorana and bosons are both massive in the expanded Hilbert space, as is directly evident from the mean field decomposition. Projecting onto the physical Hilbert space only removes states from the expanded Hilbert space, and therefore preserves the gapped nature of the excitations.

\section{Integer and Majorana modes in integer+gPf FTIs}\label{app:IntegerToMajorana}
Here we will show that if we allow for  breaking of $s_z$ conservation, then a integer+gPf FTI with $M$ Majorana edge modes, and $N$ integer modes is equivalent to a gPF FTI with $M + 2N$ Majorana edge modes. Here we will only consider the case where $M = N = 1$ integer+gPf FTI, as the other cases can be understood inductively. 

The edge Lagrangian for the $M = N = 1$ integer+gPf FTI edge takes the same form as in Eq.~\ref{eq:AppBosonLagrangian}, with $\bm{K} = \text{diag}(2,1,-2,-1)$, $\bm{\Phi} = (\phi_{\uparrow P}, \phi_{\uparrow 1}, \phi_{\downarrow P}, \phi_{\downarrow 1} )$, $\bm{t} = (1,1,1,1)$. Since we are breaking $s^z$ conservation, we will not include the spin vector $\bm{s}$. To proceed we transform the bosons as 
\begin{equation}\begin{split}
\bm{\Phi}' = \begin{pmatrix} \phi'_{\uparrow P} \\ \phi'_{\downarrow P}\\ \phi'_{\uparrow \sigma } \\ \phi'_{\downarrow \sigma}\end{pmatrix} = \bm{W} \bm{\Phi} \equiv \begin{bmatrix}  1 & 0 & 0 & 1\\ 0 & 1 & 1 & 0\\2 & 0 & 0& 1\\ 0&  2 & 1 & 0\end{bmatrix} \begin{pmatrix} \phi_{\uparrow P} \\ \phi_{I \uparrow} \\ \phi_{\downarrow P} \\ \phi_{I \downarrow}\end{pmatrix}.
\end{split}\end{equation}
Since $\bm{W}$ is a integer valued Matrix with $\det (\bm{W}) = 1$, this corresponds to an exact rewritting of the bosonic part of the Lagrangian. The new K-matrix and charge vector are $\bm{K}' =  \text{diag}(-2,2,1,-1)$, $\bm{t}' = (1,1,0,0)$. The vertex operators $\exp(i \phi_{\uparrow \sigma})$ and $\exp(-i \phi_{\downarrow \sigma})$ are charge nuetral fermions that be decomposed into two new Majoranas each, $\exp(i \phi_{\uparrow \sigma}) = \psi_{\uparrow 2} + i \psi_{\uparrow 3}$ and $\exp(-i \phi_{\downarrow \sigma}) = \psi_{\downarrow 2} - i \psi_{\downarrow 3}$. The $M = N = 1$ integer+gPf FTI and $M = 3$ gPf FTI therefore have equivalent anyons. 

To confirm that the $M = N = 1$ integer+gPf FTI and $M = 3$ gPf FTI are equivalent we also need to show that the theories have equivalent local operators. Here, local operators must have a vertex part that is written as $\exp(i \bm{l}'^T \bm{K}' \bm{\Phi}')$, and must be invariant under the parity transformation. In the original basis the parity transformation is $\psi_{\uparrow 1} \rightarrow -\psi_{\uparrow 1}$, $\phi_{\uparrow P}\rightarrow \phi_{\uparrow P} + \frac{\pi}{2}$, and similar for the spin-down operators. In the transformed basis, the parity transformations become
\begin{equation}\begin{split}
&\psi_{\uparrow 1} \rightarrow -\psi_{\uparrow 1}, \phantom{==} \phi'_{\uparrow P}  \rightarrow \phi_{\uparrow P} + \frac{\pi}{2}, \\ &\phi'_{\uparrow \sigma}  \rightarrow \phi_{\uparrow \sigma} + \pi
\end{split}\end{equation}
and similarly for the spin-down operators. Based on this, we have the following fermionic operators,
\begin{equation}\begin{split}
    &\chi^\dagger_{\uparrow 1}  = \psi_{\uparrow 1} e^{i \phi'_{\uparrow P}}, \phantom{==} 
    \chi^\dagger_{\downarrow 1}  = \psi_{\downarrow 1} e^{-i \phi'_{\downarrow P}}\\ 
    &\chi^\dagger_{\uparrow 2}  = \cos(\phi_{\uparrow \sigma}) e^{i \phi'_{\uparrow P}} =  \psi_{\uparrow 2} e^{i \phi'_{\uparrow P}}, \\
    &\chi^\dagger_{\downarrow 2}  = \cos(\phi_{\downarrow \sigma}) e^{-i \phi'_{\downarrow P}} =  \psi_{\downarrow 2} e^{-i \phi'_{\downarrow P}}, \\
    &\chi^\dagger_{\uparrow 3}  = \sin(\phi_{\uparrow \sigma}) e^{i \phi'_{\uparrow P}} = \psi_{\uparrow 3} e^{i \phi'_{\uparrow P}},\\
    &\chi^\dagger_{\downarrow 3}  = \sin(\phi_{\downarrow \sigma}) e^{-i \phi'_{\downarrow P}} = \psi_{\downarrow 3} e^{-i \phi'_{\downarrow P}}.
\end{split}\end{equation}
These are the same local operators as in the $M = 3$ gPf FTI edge. From this we conclude that the $M = N = 1$ integer+gPf FTI and $M = 3$ gPf FTI are equivalent.

\end{document}